 \documentclass[pra,amsmath,amssymb,twocolumn, showpacs, superscriptaddress,10pt]{revtex4-1}

\usepackage{amsmath}
\usepackage{hyperref}
\usepackage{graphicx}
\usepackage{amsfonts}
\usepackage{amsthm}
\usepackage{cases}
\usepackage{bm}
\usepackage{wasysym}

\usepackage[normalem]{ulem}

\usepackage{color}
\definecolor{Blue}{rgb}{0.00, 0.00, 1.00}
\definecolor{Red}{rgb}{1.00, 0.00, 0.00}

\hypersetup{
    colorlinks=true,       % false: boxed links; true: colored links
    linkcolor=red,          % color of internal links (change box color with linkbordercolor)
    citecolor=blue,        % color of links to bibliography
    filecolor=magenta,      % color of file links
    urlcolor=cyan           % color of external links
}

\newcommand{\nn}{\nonumber}
\newcommand{\be}{\begin{equation}}
\newcommand{\ee}{\end{equation}}
\newcommand{\bea}{\begin{eqnarray}}
\newcommand{\eea}{\end{eqnarray}}

\newcommand{\beq}{\begin{equation}}
\newcommand{\eeq}{\end{equation}}
\newcommand{\beqn}{\begin{eqnarray}}
\newcommand{\eeqn}{\end{eqnarray}}
\newcommand{\x}{{x}}
\newcommand{\y}{{y}}

\newcommand{\ket}[1]{\left|#1\right\rangle}
\newcommand{\bra}[1]{\left\langle #1\right|}

\newcommand{\antiquad}{\!\!\!\!\!\!\!\!}

\begin{document}

\title{Noninteracting trapped Fermions in double-well potentials: inverted parabola kernel}

\author{Naftali R. \surname{Smith}}
\affiliation{Universit{\'e} Paris-Saclay, CNRS, LPTMS, 91405, Orsay, France}
\author{David S. \surname{Dean}}
\affiliation{Univ. Bordeaux and CNRS, Laboratoire Ondes et Mati\`ere  d'Aquitaine
(LOMA), UMR 5798, F-33400 Talence, France}
\author{Pierre Le Doussal}
\affiliation{CNRS-Laboratoire de Physique Th\'eorique de l'Ecole Normale Sup\'erieure, 24 rue Lhomond, 75231 Paris Cedex, France}
\author{Satya N. \surname{Majumdar}}
\affiliation{Universit{\'e} Paris-Saclay, CNRS, LPTMS, 91405, Orsay, France}
\author{Gr\'egory \surname{Schehr}}
\affiliation{Universit{\'e} Paris-Saclay, CNRS, LPTMS, 91405, Orsay, France}

\date{\today}

\begin{abstract}

We study a system of $N$ noninteracting spinless Fermions in a confining, double-well potential in one dimension. When the Fermi energy is close to the value of the potential at its local maximum 
%(which we choose to be at $x=0$), 
we show that physical properties, such as the average density and the fermion position correlation functions, display a universal behavior that depends only on the local properties of the potential near its maximum. This behavior describes the merging of two Fermi gases, which are disjoint at sufficiently low Fermi energies. We describe this behavior in terms of a new correlation kernel that we compute analytically and we call it the ``inverted parabola kernel''. As an application, we calculate the mean and variance of the number of particles in an interval of size $2L$ centered around the position of the local maximum, for sufficiently small $L$.
Finally, we discuss the possibility of observing our results in experiments, as well as the extensions to nonzero temperature and to higher space dimensions.

\end{abstract}

\maketitle

\section{Introduction}

While Fermi gases in translationally invariant systems have been studied for a very long time, {there has been a recent surge in interest trapped Fermi gases} ~\cite{GPS08}, motivated
by cold atom experiments ~\cite{BDZ08}. The presence of a confining trap induces an edge to the Fermi gas where the density vanishes and quantum and thermal fluctuations are enhanced and play an important role.
For a large number $N$ of fermions one expects universal behavior to emerge both in the bulk as well as near the edges of this Fermi gas. Characterizing the possible universality classes of these density correlations, in particular near the edges, is an outstanding problem, {that has seen much recent activity and progress}. 

In experiments, the interactions between the fermions can be tuned using Feshbach resonances,
and it turns out that even the noninteracting limit is interesting, due to the Pauli exclusion principle. In addition it is possible to image the position of the individual atoms with high precision using the quantum Fermi microscopes \cite{Cheuk:2015,Haller:2015,Parsons:2015,Omram:2015}.
The non-interacting nature makes the many body problem analytically tractable.
In the ground state of this non interacting system, the fermion positions form a determinantal point process, which simply means that the $n$-point position correlation function can be expressed as a $n \times n$ determinant, whose entries are given by a central object called
the kernel \cite{fermions_review}. For such processes it suffices to specify the spatial behavior of the kernel to characterize 
quantum correlations. In the bulk, where the density is uniform on the scale of the inter-particle spacing, this scaled kernel
is known as the sine-kernel, which is universal, i.e. independent of the precise shape of the
trapping potential \cite{castin,CMV2011}. Near the edge several universality classes have recently been revealed,
depending on some general characteristics of the trapping potential \cite{Kohn,Joh07,Wiegmann,Eis2013,us_finiteT,fermions_review,DPMS:2015,farthest_f,LLMS17,LLMS18,WignerFunctionPaper2018,Cunden1D,Liechty17,periodic_airy,CundenAlpha} (for a recent review see \cite{FermionsRMTReview2019}). At a generic edge such that the potential is smooth near the edge, e.g. the harmonic trap, the kernel, suitably centered and scaled, converges for large $N$ to the so-called Airy kernel. Both these kernels (sine and Airy kernels) appear in random matrix theory (RMT) to describe the correlations between the eigenvalues, respectively in the
bulk and at the edge of the Wigner semi-circle describing the average density of eigenvalues
\cite{mehta,forrester,J05,Bo11}.

%In particular, for the harmonic trap, one can establish an exact correspondence between 
%the squared many body ground state wave function, $\left|\Psi_{0}\left(x_{1},\ldots,x_{N}\right)\right|^{2}$,
%and the joint PDF of the eigenvalues of a $N \times N$ random matrix drawn from the 
%Gaussian Unitary Ensemble (GUE). In this mapping the positions of the non-interacting fermions in the ground state
%are in one-to-one correspondence with the eigenvalues of the GUE random matrix \cite{CMV2011,Eis2013,marino_prl,fermions_review}..

The case of the harmonic trap played a fundamental role because it is solvable and makes an important connection between trapped noninteracting fermions and the eigenvalues of a random matrix ~\cite{FermionsRMTReview2019}. Indeed in one-dimension ($d=1$) and at zero temperature $T=0$, the squared many body ground state wave function, $\left|\Psi_{0}\left(x_{1},\ldots,x_{N}\right)\right|^{2}$, was shown to be identical to the joint probability density function (PDF) of the eigenvalues of a Gaussian Unitary Ensemble (GUE) of Random Matrix Theory (RMT)
~\cite{FermionsRMTReview2019,marino_prl,fermions_review}. Consequently, 
the positions of the fermions are in one-to-one correspondence with the eigenvalues of the GUE matrix. 
This correspondence proved to be very useful as many results from RMT could be directly
used for the trapped fermions~\cite{CMV2011,Eis2013,marino_prl,fermions_review}.
More generally, it was shown that for any smooth potential, the fermions at the edge are characterized by the Airy kernel which describes the so-called
soft edge behavior of eigenvalues in RMT \cite{fermions_review}.

Going beyond the smooth potentials, new universality classes for the edge behavior have emerged, for different types of potentials. For instance, for a infinite hard wall, the scaled kernel at the edge has been shown \citep{LLMS17,LLMS18} to be a reflected sine kernel, related to another ensemble of random matrices, known as Jacobi unitary ensemble \cite{forrester}. Another example is the inverse square barrier, which leads to the Bessel kernel \citep{Kanzieper1998, LLMS18}, which appears
in the Laguerre ensemble of RMT \cite{For93,TWBessel}. Both examples are part of a larger class of random matrix models, known as the hard edge universality class. Other interesting universality classes have also been found recently for the edge in momentum space for fermions in anharmonic potentials \citep{MulticriticalFermions2018}.

An interesting question is whether there are new universal edge behaviors when two disjoint Fermi gases are made to meet at a given point in space by tuning an external parameter. In that case {the} two edges annihilate each other, and one may anticipate a new type of scaling behavior of the quantum correlations near the merging
point. To address this question, in this paper we consider a non-interacting Fermi gas in a double well potential. By tuning the Fermi energy, or the height of the potential, one can achieve this merging scenario
(see Figure \ref{fig:potential}) and analyze the scaling behavior of the correlation functions in the large $N$ limit and in the vicinity of the merging point. Indeed, our results show that a new scaled kernel describes these correlations. We call this new kernel the {\it inverted parabola} kernel.

The merging of the supports of the fermion density is reminiscent of similar merging of
eigenvalue densities in matrix models. One well studied example is the matrix
model with one source, which leads to the so-called Pearcey kernel at the merging point
\citep{BrezinHikami1998,TWPearcey,AdlerMoerbeke2006,BleherKuijlaars2007}.
It is however different from the fermion problem in the double well potential 
studied in this paper. Indeed for the fermion problem, we show that at the merging point the density vanishes linearly, which is not the case in the random matrix problem, where the eigenvalue density vanishes with
an exponent $1/3$ \citep{BrezinHikami1998,TWPearcey}.
Hence the kernel found in this paper describes a new universality class.

Even though in this paper we focus on $d=1$ and zero temperature, our results can be extended
to higher dimensions and finite temperature, as has been done recently for the other edge universality classes discussed above.

The paper is organized as follows: in Section \ref{sec:2} we define the model, and discuss the important length and energy scales which control the physics. In
order to make the paper self contained, in Section~\ref{sec:DPP} we give a brief reminder of the determinantal structure of the spatial correlation functions in terms of a central object called the kernel. We also present two different methods that are used to calculate the kernel.
In section~\ref{sec:denisty_and_kernel} we calculate the density and the kernel {near the local maximum of the potential, for Fermi energies which are very close to the value of the potential there.} 
%\sout{at $x\simeq0$ and $\mu \simeq V_0$.}
We show that these become universal in the large-$N$ limit.
%, and depend only on the second derivative of the potential at $x=0$ (and not on higher derivatives).
As an application of these results, in Section~\ref{sec:number_variance} we calculate the {mean and} the variance of the number of fermions in the interval $x\in\left[-L,L\right]$ around the local maximum. 
%\sout{and the ``hole probability'' - the probability that no particles are found in this interval}.
Finally, we discuss our results and several
extensions in section~\ref{sec:discussion}.
Some of the technical details are relegated to the Appendices.

% \cite{LLMS17,LLMS18, FermionsRMTReview2019}.
%\cite{DPMS:2015,farthest_f} 
%\cite{fermions_review, FermionsRMTReview2019}.
%%
%\cite{BB91,For93}
%\cite{us_finiteT,fermions_review} \cite{fermions_review}. 
%
%\citep{BrezinHikami1998}. 
% \citep{BrezinHikami1998} 

\section{The model, the setup, and the scales} 
\label{sec:2} 

We consider $N$ noninteracting spinless fermions, each of mass $m$, in one dimension $d=1$.
The $N$ body Hamiltonian is $\hat {\cal H}_N=\sum_{i=1}^N \hat H(\hat x_i,\hat p_i)$, where the 
single particle Hamiltonian is given by
\be
\hat H = \hat H(\hat x,\hat p)=\frac{\hat p^2}{2 m} + V(\hat x) \;.\label{ham} 
\ee
%with $V\left(\x\right)=\frac{1}{2} m \omega^2 \x^2$ in the case of the harmonic oscillator. 
Let $\Psi_{0}\left(x_{1},\ldots,x_{N}\right)$ denote the ground state many body wave function 
of ${\cal H}_N$. The quantum probability density in the ground state is given by the squared wave function
$\left|\Psi_{0}\left(x_{1},\ldots,x_{N}\right)\right|^{2}$. The ground state is constructed as a Slater determinant involving the first $N$ single particle energy levels with one fermion at each level. The energy of the highest occupied single particle level is the Fermi energy denoted by $\mu$, which is an increasing function of~$N$.

%\begin{figure}[ht]
%\includegraphics[width = 0.8 \linewidth]{potential.pdf}
%\caption{
%Solid line: the double-well potential $V\left(x\right)$. Dashed line: the quadratic approximation~(\ref{eq:quadratic_potential}).} \label{fig:potential}
%\end{figure}

\begin{figure*}[ht]
	\includegraphics[width = 0.32 \linewidth]{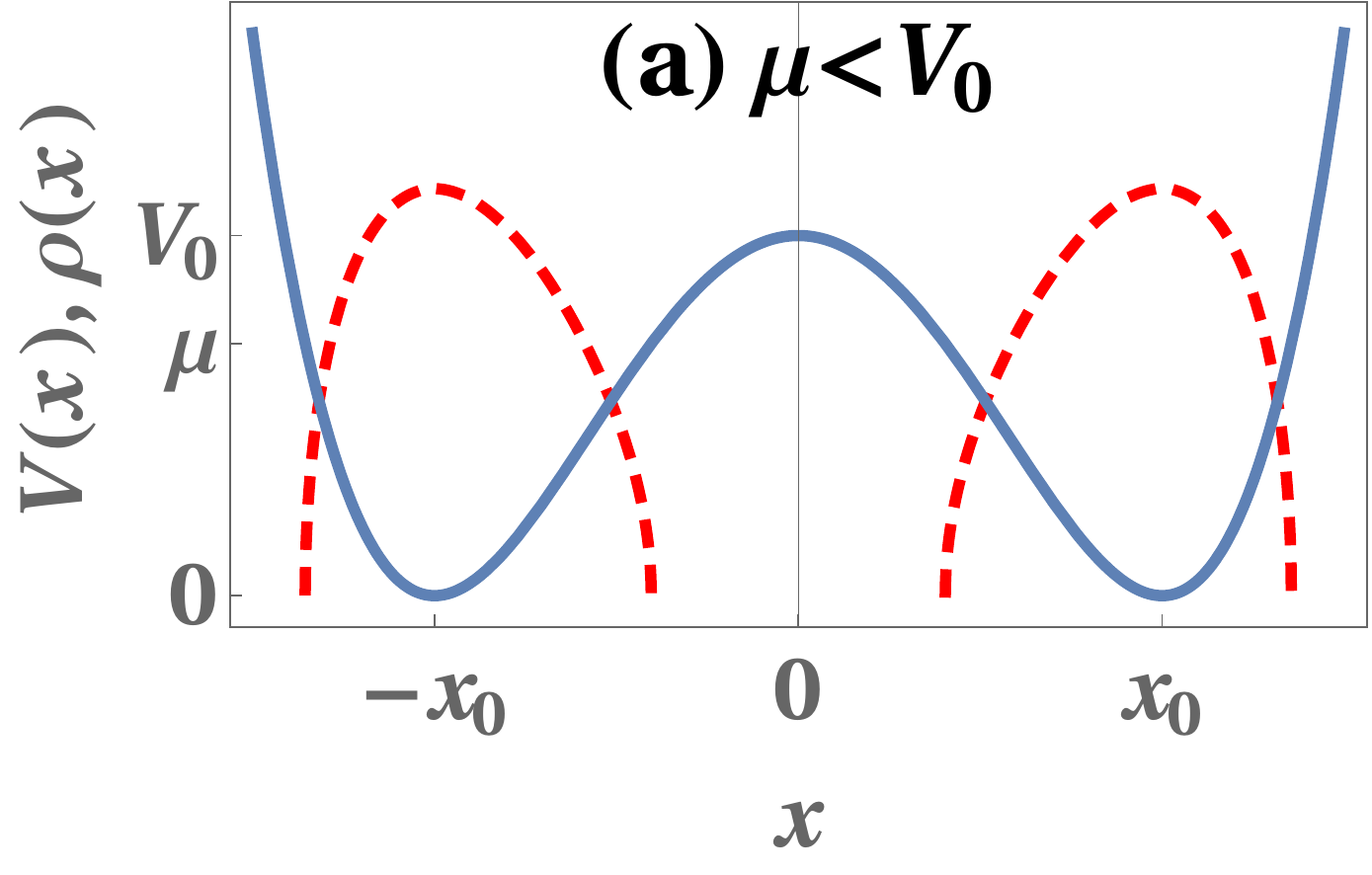}
	\includegraphics[width = 0.32 \linewidth]{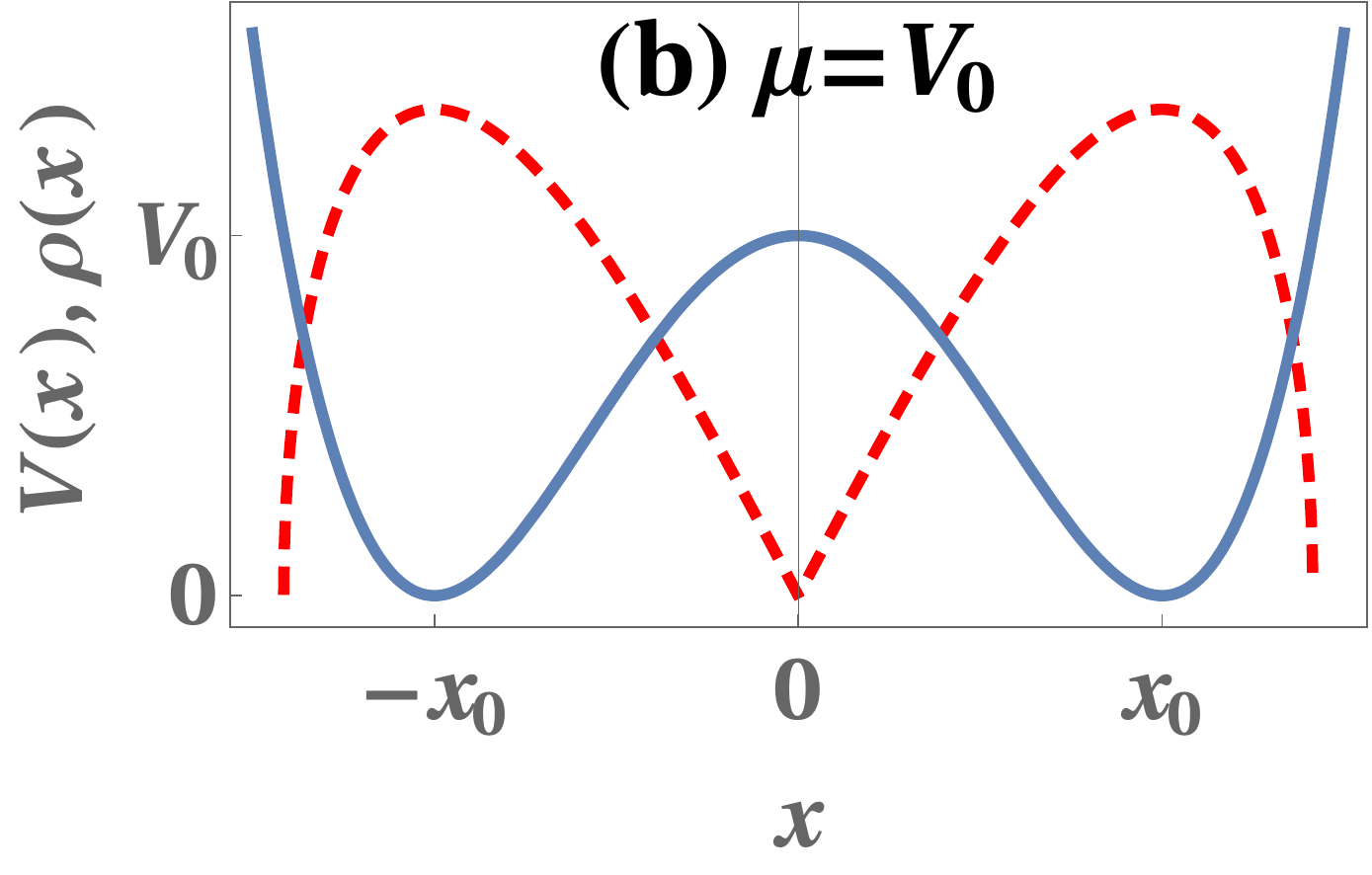}
	\includegraphics[width = 0.32 \linewidth]{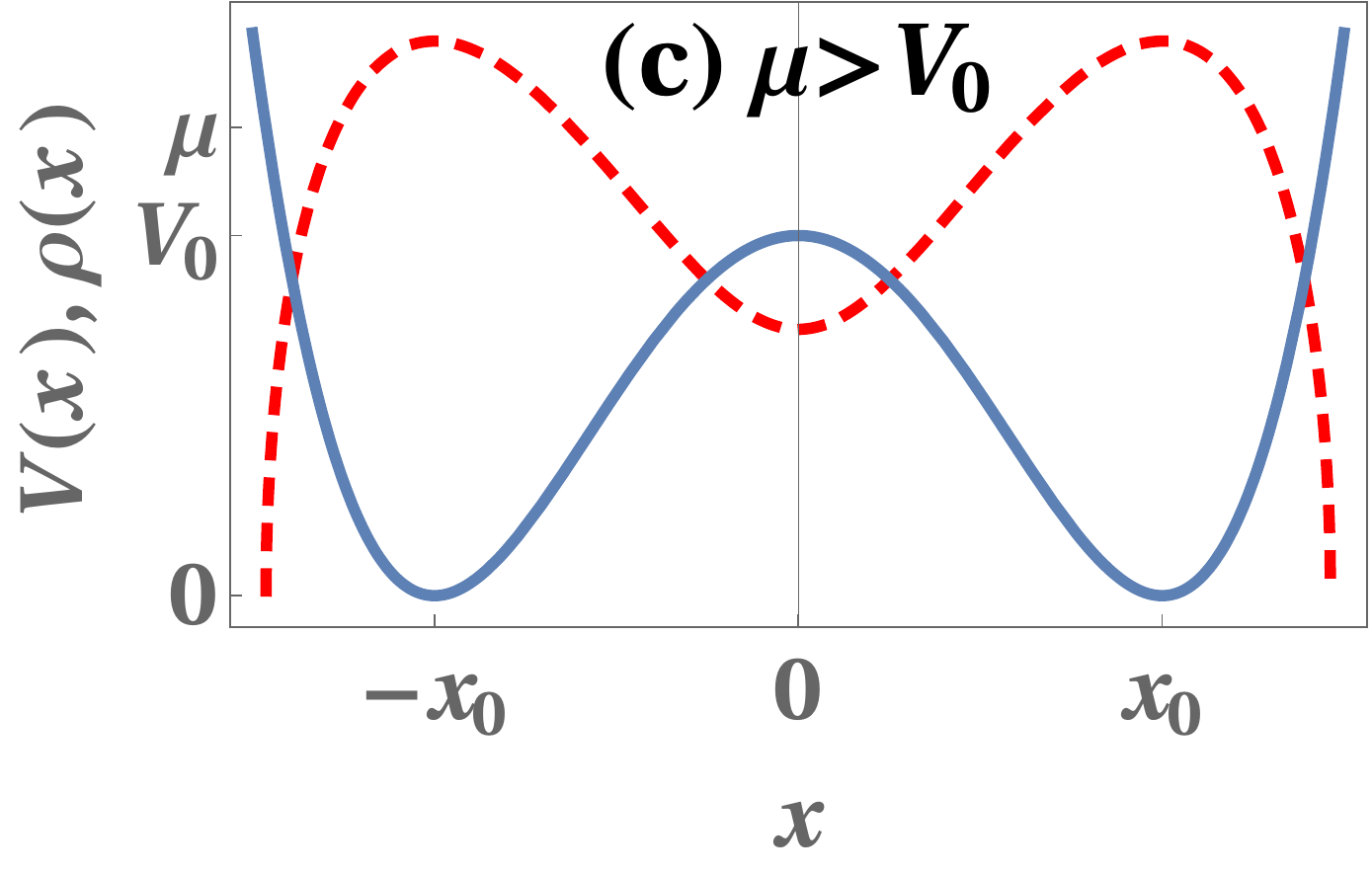}
	\caption{Solid line: the double-well potential $V\left(x\right)=V_{0}v\left(x/x_{0}\right)$. Dashed red line: associated average density $\rho^{\rm bulk}(x)$ from Eq.~\eqref{eq:bulk_density},
	in subcritical $\mu<V_0$ (a), critical $\mu=V_0$ (b) and supercritical $\mu>V_0$ (c) regimes. In the subcritical regime there are two disjoint supports for the density, which merge at the critical point where the density vanishes linearly at $x=0$. In the supercritical case there is a single support with a local minimum of the density at $x=0$.} 
	\label{fig:potential}
\end{figure*}

In this paper we are interested in double-well potentials. We assume that the potential has unique energy and length scales $V_{0}$ and $x_0$ respectively, that is, it is of the form
\beq
\label{eq:potentialScaling}
V\left(x\right)=V_{0} \, v\left(\frac{x}{x_{0}}\right)
\eeq
where $v\left(z\right)$ is dimensionless, with a simple double-well form, see Fig.~\ref{fig:potential},
with a smooth parabolic behavior near its single local maximum $v\left(0\right)=1$ [so that $v'\left(0\right) = 0$ and $v''\left(0\right)<0$]. The potential is confining $v\left(x\to\pm\infty\right)\to\infty$, in order to trap the fermions.
{As a concrete example, one can consider $v\left(z\right)=\left(z^{2}-1\right)^{2}$.}

Since we are interested in the large $N$ limit, equivalently large $\mu$, we need to scale the parameters of the potential with $\mu$. In order to probe the vicinity of the region around the local maximum of the potential at $x=0$, we clearly need to scale $V_0 = O(\mu)$. The scaling of $x_0$ is in principle arbitrary and may depend on the experimental setup, however we choose here to scale it as $x_{0}=O\left(\sqrt{\mu}\right)$. In this case,
as we show below, see Eq.~\eqref{eq:N0def}, $N=O\left(\mu\right)$ as in the case of the harmonic trap. To summarize we choose
\be \label{scale} 
V_0 = O\left(\mu\right) \quad , \quad x_{0}=O\left(\sqrt{\mu}\right) \quad , \quad N = O\left(\mu\right) \;.
\ee 

One of the central observables is the 
average density of fermions $\rho_{N} \! \left(x\right)$, normalized to unity (and not to the total number of fermions) and defined by
\be
\rho_{N}\left(x\right)=\frac{1}{N}\left\langle \sum_{i=1}^{N}\delta\left(x-x_{i}\right)\right\rangle _{0}
\ee
where $\langle \dots \rangle_0$ denotes expectation values w.r.t. the ground state. The typical inter-particle spacing is, {locally, given by} $\propto1/\rho_{N}\left(\x\right)$.
In the large $N$ limit it is well known that the average density in the bulk of the Fermi gas 
is well described \citep{fermions_review, WignerFunctionPaper2018} by the local density approximation (LDA) \citep{castin}, $\rho_N(x) \simeq \rho^{\rm bulk}\left(\x\right)$ with
\beq
\label{eq:bulk_density}
N \rho^{\rm bulk}\left(\x\right)=\frac{\left[2 m\left(\mu-V\left(\x\right)\right)\right]^{1/2}}{\pi \hbar}\Theta\left[\mu-V\left(\x\right)\right]
\eeq
where $\Theta\left(z\right)$ is the Heaviside function. The Fermi energy is related to the total number of particles
$N$ as follows
\bea
\label{eq:NasFunctionOfMu}
\antiquad N &=&\int_{-\infty}^{\infty} N\rho_N\left(x\right)dx\nonumber\\
&\simeq & \int_{-\infty}^{\infty}dx\,\Theta\left[\mu-V\left(x\right)\right]\frac{\left[2m\left(\mu-V\left(x\right)\right)\right]^{1/2}}{\pi\hbar}.
\eea
This equation determines $\mu$ as a function of $N$ for 
a given potential $V(x)$. 

The estimate
$\rho_N(x) \simeq \rho^{\rm bulk}(x)$ breaks down 
however in regions where the density becomes small.
%i.e. $V(x) - \mu$ small (compared to $\mu$).
For the double well potential, if $\mu<V_0$ the bulk density has two disjoint supports
with four edges. For $\mu>V_0$ the bulk density has a single support with two edges.
The most interesting case is when $\mu$ approaches $V_0$ from below, when the two
disjoint {supports} merge (see Figure \ref{fig:potential}). Exactly at $\mu=V_0$ the bulk density has three edges, two outer
edges and one inner edge at $x=0$, see Figure \ref{fig:density}. We are particularly interested
in this inner edge regime, where we show that new physics emerges.

Focusing near $x=0$, and at $\mu=V_0$, let us first estimate the width $w_N$ of this 
inner edge regime, where the bulk density breaks down. This is done by setting 
\be \label{width0} 
\int_{-w_{N}}^{w_{N}}N\rho^{\rm bulk}\left(x\right)dx\sim1
\ee 
which means that the typical number of particles in the interval $[-w_N,w_N]$ is
of order unity. Near its local maximum at $x=0$ the double well potential can be well approximated 
by the inverted parabolic form 
\beq
\label{eq:quadratic_potential}
V\left(x\right)\simeq V_{0}-V_{2}x^{2},\qquad\left|x\right|\ll x_{0} 
\eeq
where
\beq
\label{eq:V2def}
V_{2}=-\frac{1}{2}V''\left(0\right)=-\frac{V_{0}}{2x_{0}^{2}}v''\left(0\right) \,> 0.
\eeq
Plugging the approximate form of the potential near $x=0$, Eq.~(\ref{eq:quadratic_potential}), into the formula for the bulk density (\ref{eq:bulk_density}), we obtain the bulk density in the vicinity of $x=0$ as
\beq
\label{eq:bulk_denisity_DW}
\! N \rho^{\rm bulk}\!\left(x\right) \! \simeq \! \frac{\left[2m\left(\mu-V_{0}+V_{2}x^{2}\right)\right]^{1/2} \!  \Theta \! \left(\mu-V_{0}+V_{2}x^{2}\right)}{\pi\hbar}.
\eeq
At $\mu=V_0$ one has $N\rho^{\rm bulk}\!\left(x\right) \sim |x|$ near $x=0$, hence the bulk
density vanishes linearly at $x=0$. This is in contrast with the density vanishing as $\sqrt{x_e-x}$, {where $x_e$ is the location of the edge $V\left(x_{e}\right)=\mu$,}
at the outer edges, as in the 
Wigner semi-circle for the harmonic trap. This linear behavior of the density
gives rise to a different universality class and a new kernel for the quantum correlations, as
we will show below. In agreement with \eqref{width0} we define the width $w_N$ of the inner edge around $x=0$ to be
\beq
\label{eq:wNdef}
w_{N}=\frac{\hbar^{1/2}}{\left(mV_{2}\right)^{1/4}}.
\eeq
With the scaling \eqref{scale} we see that $V_2 = O\left(1\right)$ and consequently $w_N=O\left(1\right)$, in the limit of large $\mu$. 

This allows us to discuss the domain of validity of approximating the potential
$V(x)$ by its quadratic expansion in \eqref{eq:quadratic_potential}. Indeed for
$x$ close to $x=0$ on a scale $x \sim w_N$ the $n$-th order anharmonic correction
in the Taylor expansion of $V(x)$ is $\frac{1}{n!} v^{(n)}(0) V_0 (w_N/x_0)^n$,
where $v^{(n)}(0)$ is the $n$-th derivative of $v(z)$ at $z=0$.
The condition for it to be negligible compared to the second order term reads
\be
w_N^{n-2} \ll \frac{V_2}{V_0} x_0^n \quad , \quad n \geq 3
\ee
With our choice of $V_0$ and $x_0$ as in \eqref{scale} we see that this condition reduces
to $w_N = O\left(1\right) \ll \sqrt{\mu} \sim x_0$, which is realized for large $\mu$, the limit
studied here.

Alternatively, $w_N$ can be deduced from dimensional analysis: it is the only length scale that one can construct from $m$, $V_2$ and $\hbar$. 
Similarly, the only energy scale that one can construct from these quantities is $\hbar\sqrt{V_{2}/m}$, motivating us to define a shifted and rescaled dimensionless Fermi energy
\beq
\label{eq:mu1def}
\mu_{1}=\frac{\sqrt{m}\,\left(\mu-V_{0}\right)}{\hbar\sqrt{2V_{2}}}.
\eeq
As we will see, $\mu_{1}$ arises naturally in the calculations that we perform below. The factor $\sqrt{2}$ in Eq.~(\ref{eq:mu1def}) is included for later convenience.

\begin{figure}[ht]
\includegraphics[width = 0.7\linewidth, angle = 90]{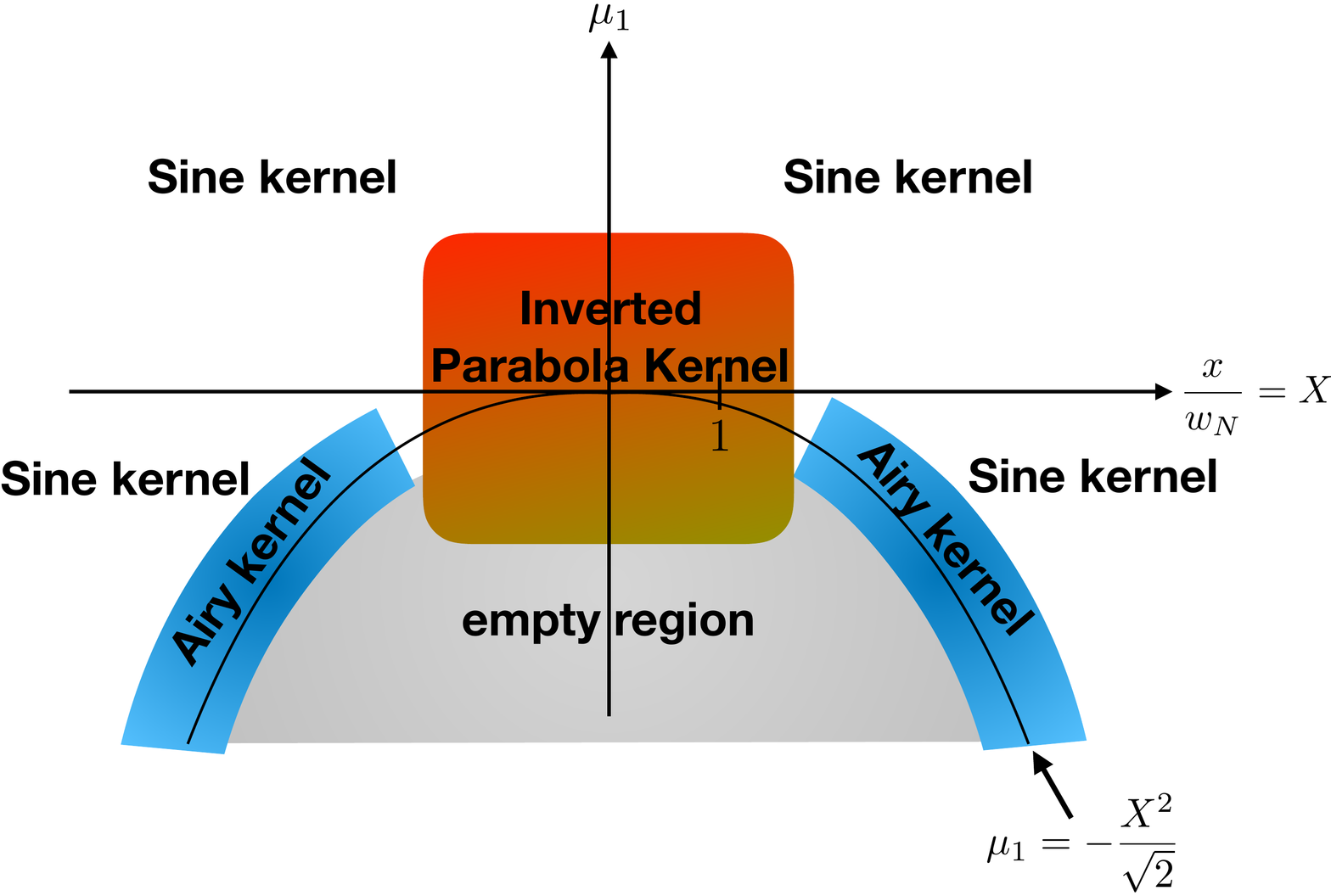}
\caption{Phase diagram of the different regimes of density and correlations of the
Fermi gas in the double well potential, shown in the plane $\left(x/w_{N},\mu_{1}\right)$, where
$w_N$ and $\mu_1$ are defined in Eqs.~\eqref{eq:wNdef} and \eqref{eq:mu1def}, respectively.
The sine-kernel region defines the bulk behavior. The blue shaded region around the 
edge of the density for $\mu_1<0$ is of spatial width $w_N^s / w_N$ and corresponds to the standard edge regime described by the Airy kernel.
The red shaded square around the center corresponds to the region where the novel behavior occurs, characterized by the inverted parabola kernel. In the lower region (shaded grey), there are no particles.} 
\label{fig:diagram}
\end{figure}

Depending on the value of $\mu_1$ one can probe different types of density and correlations in the Fermi gas
by varying the dimensionless reference point $x/w_N$. This is illustrated in the $(x/w_N,\mu_1)$ plane
in Figure \ref{fig:diagram}. There are four regions in that plane. The bulk behavior, characterized by the sine-kernel, which leads to a density given by \eqref{eq:bulk_denisity_DW},
i.e. in dimensionless form
\be \label{bd} 
N\rho_{N}\left(x\right)\simeq\frac{\sqrt{2}}{\pi w_{N}}\sqrt{\left(\frac{x}{w_{N}}\right)^{2}+\sqrt{2} \, \mu_{1}}
\ee
is valid whenever $\sqrt{2}\,\mu_{1}+\left(x/w_{N}\right)^{2}\gg1$. The standard edge behavior, characterized by the Airy kernel, holds in the region around
the curve $\mu_{1}=-2^{-1/2}\left(x/w_{N}\right)^{2}$, i.e.
around the edges $x= \pm x_e$ with $x_{e}=2^{1/4}\left|\mu_{1}\right|^{1/2} \! w_{N}$. 
The new universality class studied in this paper, characterized by the inverted parabola kernel, see below, appears in the region around the center such that $|\mu_1|=O\left(1\right)$ and $x/w_N = O\left(1\right)$.
Finally, the region $\mu_1<0$ and $|x|<x_e$ is empty of particles.

The precise behavior of the correlation functions, and the kernel, are derived in section \ref{sec:denisty_and_kernel}.
Our main results for the scaled inverted parabola kernel near the inner edge,
can be read off Eqs.~(\ref{eq:K_scaling}) and~(\ref{eq:K_IP_def}). Similarly the scaled density
near the inner edge is given in Eqs.~(\ref{eq:rho_scaling}) and~(\ref{eq:F_IP_def}) below.

\section{Determinantal structure of the spatial correlations at zero temperature}
\label{sec:DPP}

\subsection{General framework}

As mentioned in the introduction the ground state many-body wave function $\Psi_0$ can be expressed as 
an $N \times N$ Slater determinant,
\bea \label{slater} 
\Psi_{0}\left({x}_{1},\cdots,{x}_{N}\right)=\frac{1}{\sqrt{N!}}\,\det\left[\psi_{k}\left({ x}_{j}\right)\right]_{1\leq k,j\leq N} .
\eea 
Here $\psi_{k}({x})$'s are the normalized single particle eigenfunctions of the Hamiltonian \eqref{ham} (and their corresponding energy eigenvalues $\epsilon_{ k}$), labeled by the integers ${k}$.
The determinant in \eqref{slater} is constructed from the first $N$ single 
particle eigenfunctions labeled by the sequence $k=1, \ldots, N$, with 
non-decreasing energies such that $\epsilon_{N} = \mu$ where $\mu$ is the Fermi 
energy. The factor $1/ \sqrt{N!}$ ensures that the many body wave function is normalized to unity.
%For a sufficiently confining potential, $\mu$ generically increases with increasing  $N$~\cite{eisler_prl,Castin}.

The quantum spatial fluctuations are encoded in the joint PDF
\beq\label{jpdf}
P_{{\rm joint}}\left(\x_{1},\cdots,\x_{N}\right)=\left|\Psi_{0}\left(\x_{1},\cdots,\x_{N}\right)\right|^{2}.
\eeq
Of special interest are the $n$-point spatial correlation functions $R_{n}\left(\x_{1},\cdots,\x_{n}\right)$, with $1 \leq n \leq N$, which are given by the different marginals of the full joint PDF, i.e.,~\cite{mehta,forrester}
\bea \label{def_correl} 
&&\antiquad \! R_{n}\left(\x_{1},\cdots,\x_{n}\right)=\frac{N!}{\left(N-n\right)!}\nonumber\\
&&\antiquad \! \times \!\! \int \!\! d{\x}_{n+1} \cdots \!\! \int \!\! d{\x}_{N} P_{{\rm joint}} \! \left(\x_{1},\cdots,\x_{n},\x_{n+1},\cdots,\x_{N}\right) ,
\eea
where the integrals over the positions $\x_i$'s run over their full domain of definition. In particular, for $n=1$ 
\be \label{def_density}
\!\!\! R_{1}\left(\x\right)=N \int d\x_{2}\cdots\int d\x_{N}P_{{\rm joint}}\left(\x,\x_{2},\cdots,\x_{N}\right),
\ee
which is directly related to the average density of fermions in the ground-state via
\be \label{rel_R1_density}
\!\!\! R_{1}\! \left(\x\right)=N \! \rho_{N} \! \left(\x\right) ,\quad \rho_{N} \! \left(\x\right)=\frac{1}{N}\left\langle \sum_{i=1}^{N}\delta\left(\x-\x_{i}\right) \! \right\rangle _{0} ,
\ee
where $\langle \cdots \rangle_0$ denotes an average in the ground state~(\ref{slater}). 
%Note that this density $\rho_N\left(\x\right)$ is normalized to unity, and not to the total number of fermions.  
One can show (see e.g. \cite{fermions_review}) that for any $1 \leq n \leq N$, the $n$-point correlation function $R_{n}\left(\x_{1},\cdots,\x_{n}\right)$ can be written as an $n \times n$ determinant constructed from a central object, the so-called kernel \cite{mehta,forrester,fermions_review} 
\bea
\label{determinantal_structure}
R_{n}\left(\x_{1},\cdots,\x_{n}\right)=\det_{1\leq\,j,\,l\,\leq n}K_{\mu}\left(\x_{j},\x_{l}\right)\;,
\eea
with
\be 
K_{\mu}\left(\x,\x'\right)=\sum_{{ k}}\Theta\left(\mu-\epsilon_{{ k}}\right)\psi_{{ k}}^{*}\left(\x\right)\psi_{{ k}}\left(\x'\right).\label{Kmu}
\ee
The above properties rely, in great part, on the fact that the kernel has the important property of being reproducible, that is to say
\begin{equation}
\int d\y \ K_{\mu}\left(\x,\y\right)K_{\mu}\left(\y,\x'\right)= K_{\mu}\left(\x,\x'\right).\label{reprod}
\end{equation}
%is defined in terms of the single particle eigenfunctions $\psi_{ k}\left(\x\right)$ of \eqref{ham}
%and their associated eigenenergies $\epsilon_{ k}$, labeled by quantum numbers ${ k}$. 
%In \eqref{Kmu} $\mu$ is chosen so that the sum contains exactly $N$ levels. 
The property~(\ref{determinantal_structure}) establishes that the positions of $N$ noninteracting fermions trapped in an arbitrary potential $V\left(x\right)$ constitute a determinantal point process \cite{J05, Bo11} with a kernel given by Eq. (\ref{Kmu}).  
In particular, for $n=1$, this result (\ref{determinantal_structure}), together with the relation in (\ref{rel_R1_density}), implies
\beq
\label{coinciding}
N\rho_{N}\left(\x\right)=K_{\mu}\left(\x,\x\right)=\sum_{{ k}}\Theta\left(\mu-\epsilon_{{ k}}\right)\left|\psi_{{ k}}\left(\x\right)\right|^{2}.
\eeq

We end this section with a brief description of an alternative method for expressing the kernel \citep{fermions_review}. This method makes use of the Euclidean propagator associated to the one body  Hamiltonian~(\ref{ham}). By definition, it obeys the imaginary time 
 Schr\"odinger equation
 \begin{equation}
 -\hbar{\partial G({ x},{ y};t)\over \partial t}=  \hat H G({ x},{ y};t),\label{itshro}
 \end{equation}
 where $\hat{H}=\hat{H}\left({ y},\frac{\hbar}{i}\nabla_{{ y}}\right)$ is the quantum Hamiltonian~(\ref{ham}) acting on the variable ${ y}$ (in our convention), with the initial condition
 \begin{equation}
G\left({ x},{ y};0\right)=\delta \left({ x}-{ y}\right).
 \end{equation}
Its expression as a function of the eigenstates is given by 
\begin{eqnarray}
\label{eq:Gdef}
G\left({ x},{ y};t\right)  = \sum_{{ k}}\psi_{{ k}}^{*}\left({ x}\right)\psi_{{ k}}\left({ y}\right)\exp\left(-\frac{\epsilon_{{ k}}t}{\hbar}\right).
 \end{eqnarray}
Taking a derivative of Eq. \eqref{Kmu} with respect to $\mu$, followed by a Laplace transform with respect to $\mu$, leads to
\begin{eqnarray}
\label{eq:Gdef}
G\left({ x},{ y};t\right)  &=& \int_0^\infty d\mu \exp\left(-{t\mu \over \hbar}\right)
\partial_\mu K_\mu({ x},{ y}) \nonumber \\
 &=& {t\over \hbar}\int_0^\infty d\mu \exp\left(-{t\mu \over \hbar}\right)
K_\mu({ x},{ y}) 
 \end{eqnarray}
where in the last line we performed an integration by parts. 
Finally, the kernel is obtained from the propagator via the 
Bromwich inversion formula for Laplace transforms:
\begin{equation}
\label{laplace_inverse}
K_{\mu}\left({ x},{ y}\right)=\int_{\Gamma}\frac{dt}{2\pi it}\exp\left(\frac{\mu t}{\hbar}\right)G\left({ x},{ y};t\right)\;,
\end{equation}
where $\Gamma$ indicates the Bromwich integration contour in the complex plane \citep{fermions_review}.

\subsection{Standard bulk and edge kernels} 

In the bulk, the average local density is well approximated by $\rho_N(x) \simeq \rho^{\rm bulk}(x)$, 
where $\rho^{\rm bulk}(x)$ is given in 
Eq.~\eqref{eq:bulk_density}. The correlations are described on the scale $1/\left[N\rho_{N}\left(x\right)\right]$ 
by the celebrated sine kernel as
\be
\label{eq:sine_kernel_def}
K_{\mu}\left(x,y\right)\simeq\frac{\sin\left[k_{F}\left(x\right)\left(x-y\right)\right]}{\pi\left(x-y\right)}
\ee 
where $k_{F}\left(x\right)=\sqrt{2m\left[\mu-V\left(x\right)\right]}\,/\hbar=\pi N\rho^{{\rm bulk}}\left(x\right)$ is the local Fermi wave vector.

The average bulk density $\rho^{\rm bulk}\left(\x\right)$ exhibits an edge at $\x_{e}$, whenever $\x_{e}$ satisfies $V(\x_{e}) = \mu$, beyond which it vanishes. When the gradient of the potential at $\x_{e}$ is nonzero, $V'(x_e) \neq 0$, the bulk density $\rho^{\rm bulk}\left(\x\right)$ vanishes near the edge as $\sqrt{|x-x_e|}$. This is called a soft edge. The exact average density gets smeared over a width 
$w_{N}^{s}=\hbar^{2/3}2 m \left|V'\left(x_{e}\right)\right|^{-1/3}$ around the soft edge $x_e$
(where the superscript $s$ in $w_N^s$ denotes the soft edge).
The kernel takes the following scaling form around the edge 
(where $V'\left(x_{e}\right)>0$) \citep{fermions_review}
\be
\label{softkernel} 
K_{\mu}\left(x,y\right)\simeq\frac{1}{w_{N}^{s}}K_{{\rm Ai}}\left(\frac{x-x_{e}}{w_{N}^{s}},\frac{y-x_{e}}{w_{N}^{s}}\right)
\ee
in terms of the standard Airy kernel~\cite{mehta,forrester}
\bea
\label{airy_kernel.1}
K_{\rm Ai}(a,b) &=&
\frac{{\rm Ai}(a){\rm Ai}'(b)-{\rm Ai}'(a){\rm Ai}(b)}{a-b} \nn\\
&=&  \int_0^{+\infty} du \, {\rm Ai}(a+u) {\rm Ai}(b+u) \, .
\eea
At coinciding points the density takes the scaling form \cite{BB91,fermions_review}
\be 
\label{softdens} 
\rho_{N}\left(x\right)=\frac{1}{w_{N}^{s}}F_{1}\left(\frac{x  - x_e}{w_{N}^{s}}\right)
\ee 
where the scaling function $F_1(z)={\rm Ai}'(z)^2- z {\rm Ai}(z)^2$.
%{\blue For a ``left edge'', $V'\left(x_{e}\right)<0$, the signs of the arguments of the scaling functions $K_{\rm Ai}(a,b)$ and $F_1(z)$ in Eqs.~\eqref{softkernel} and~\eqref{softdens} respectively should be flipped.}

All the results of this section, both in the bulk as well as near the
soft edges have been extended to finite temperature and higher
space dimensions, see Refs.~\citep{us_finiteT,DPMS:2015,fermions_review,Liechty17,FermionsRMTReview2019}
for explicit formulae.
In $d=1$, for the harmonic potential, the characteristic temperature scale in the bulk is 
$T \sim \mu \sim N$
while at the edge it is 
$T \sim N^{1/3}$.

\section{Density and kernel for double-well potentials near criticality}

\label{sec:denisty_and_kernel}

In this section we compute the density and the kernel in the critical region $\mu \approx V_0$ near the local maximum of the double well potential at $x=0$, using the inverted parabola approximation
as in Eq.~(\ref{eq:quadratic_potential}). There are two alternative ways to perform the calculation, the first uses the relation between the kernel and the Euclidean propagator \eqref{laplace_inverse}. The second
method uses the definition of the kernel as a sum over eigenfunctions, as in \eqref{Kmu}. The two methods
lead to two rather different looking expressions, which however can be verified to be equivalent.

\subsection{Representation in terms of the propagator}

\begin{figure}[ht]
	\includegraphics[width = 0.98 \linewidth]{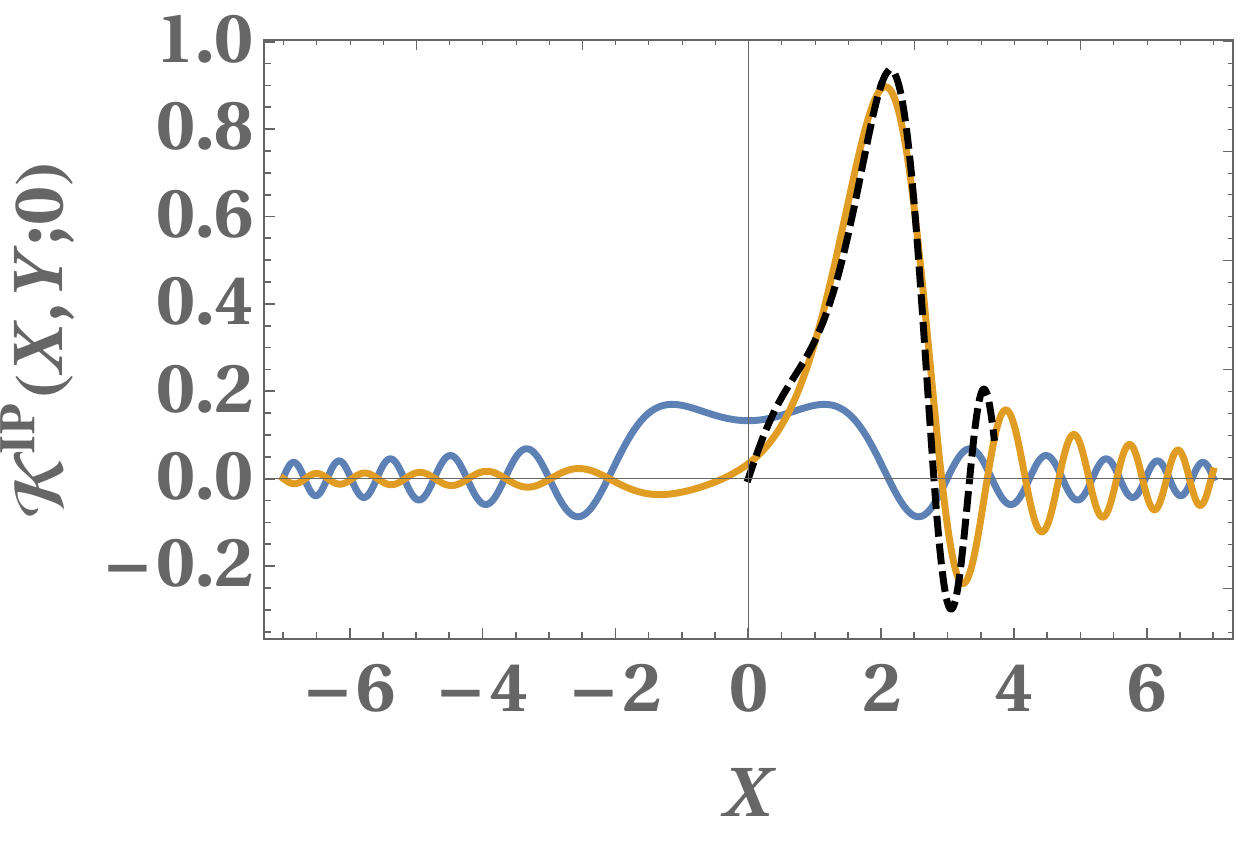}
	\caption{
		Solid: The rescaled kernel $\mathcal{K}^{IP}\left(X,Y;\mu_{1}\right)$ as a function of $X$ for $\mu_{1}=0$ and $Y=0$ (in blue) and $Y=2$ (in orange). Dashed: sine-kernel approximation \eqref{eq:K_IP_sine_kernel_approx} for $Y=2$. The approximation \eqref{eq:K_IP_sine_kernel_approx} is good at $X\simeq Y$ and $\mu_{1}+X^{2} \! /\sqrt{2}\gg1$.
		%(b) The bulk approximation.
	} \label{fig:kernel0}
\end{figure}

Our starting point is the exact relation between the kernel and the Euclidean propagator \eqref{laplace_inverse}. 

\begin{figure*}[ht]
	\includegraphics[width = 0.49 \linewidth]{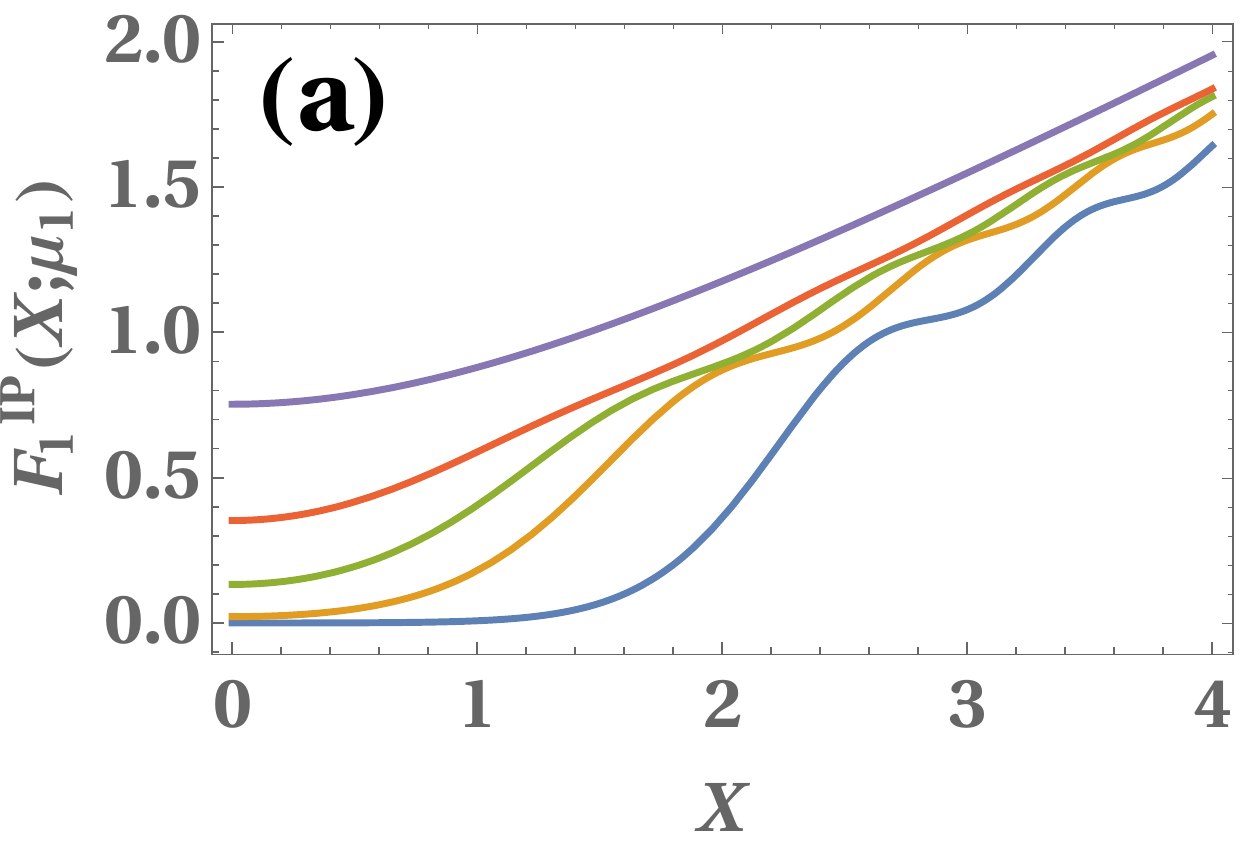}
	\includegraphics[width = 0.49 \linewidth]{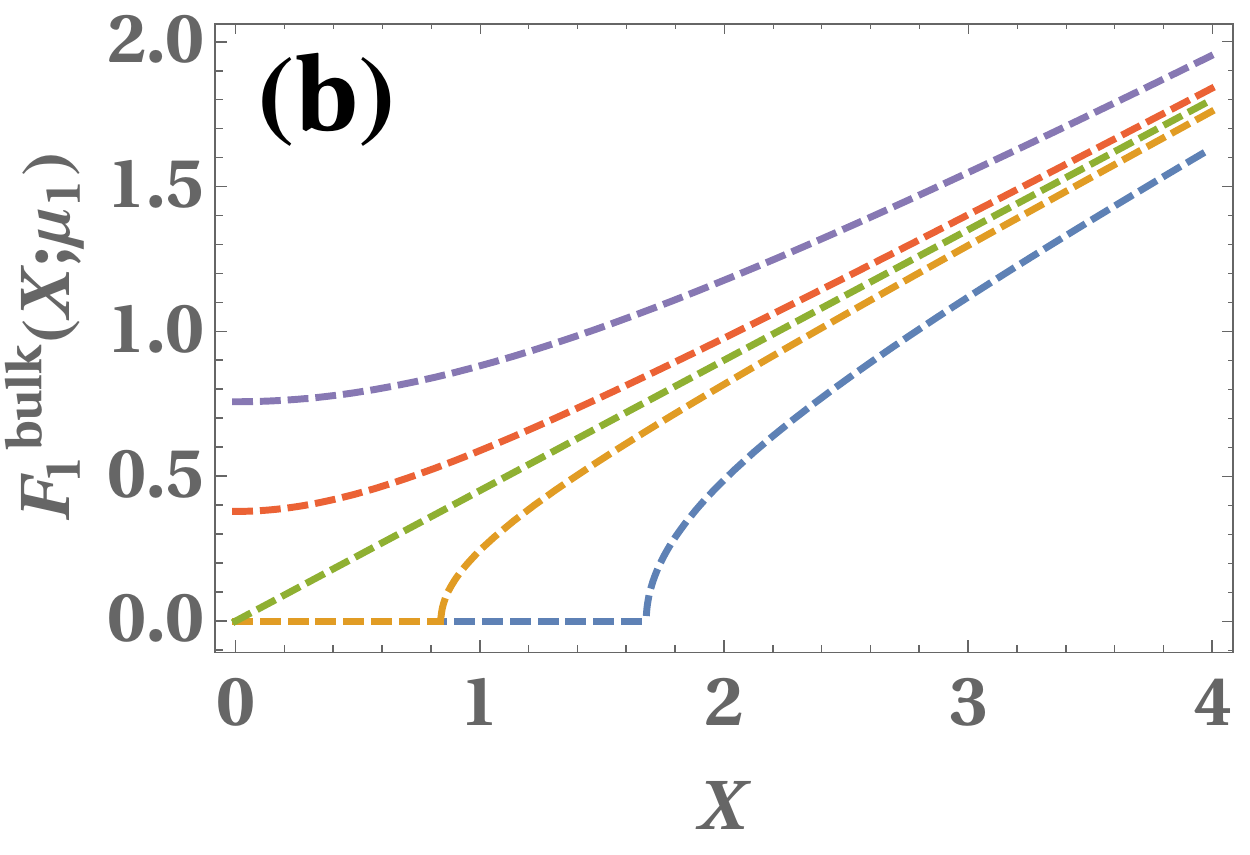}
	\caption{
%		(a) The bulk approximation (\ref{eq:F_IP_bulk_approx2}) for the
%		rescaled density $F_{1}^{IP} \! \left(X;\mu_{1}\right)$ as a function of $X$ \citep{footnote:mirror_symmetry} for $\mu_{1}=-2,-0.5,0,0.5,2$ (bottom to top).
%		(b) The exact rescaled density $F_{1}^{IP} \! \left(X;\mu_{1}\right)$ as a function of $X$, see Eq.~(\ref{eq:F_IP_def}), for the same values of $\mu_{1}$ as in (a).
		(a) The exact rescaled density $F_{1}^{IP} \! \left(X;\mu_{1}\right)$ as a function of $X$ \citep{footnote:mirror_symmetry}, see Eq.~(\ref{eq:F_IP_def}), for $\mu_{1}=-2,-0.5,0,0.5,2$ (bottom to top).
		(b) The bulk approximation (\ref{eq:F_IP_bulk_approx2}) for the
		rescaled density $F_{1}^{IP} \! \left(X;\mu_{1}\right)$ for the same values of $\mu_{1}$ as in (a).	
} 
	\label{fig:density}
\end{figure*}

The propagator $G(x,y;t)$ for the standard harmonic oscillator $V(x) = V_0 +  m \omega^2 x^2 \! / 2$
is known exactly \citep{Feynman} as
\be
\label{h}
\! G\left(x,y;t\right)=\frac{\exp\left[-\frac{V_{0}t}{\hbar}- \frac{m \omega}{2 \hbar} \frac{\left(x^{2}+y^{2}\right)\cosh\left(\omega t\right)-2xy}{\sinh\left(\omega t\right)}\right]}{\left[2 \pi \hbar \sinh\left(\omega t\right)/(m \omega) \right]^{1/2}}.
\ee
The propagator for the inverted harmonic oscillator with potential 
$V(x)=V_0- V_2 x^2$ can be obtained from \eqref{h} 
by setting $\omega = i \nu$ where $\nu=\sqrt{2V_{2}/m}$, and 
one obtains 
%\bea
%\label{eq:propagatorIP}
%G\left(x,y;t\right)&=&\frac{1}{\left[\sqrt{2}\,\pi\sin\left(\nu t\right)\right]^{1/2}w_{N}} \nn\\
%&\times&\exp\left[-\frac{V_{0}t}{\hbar} -\frac{\left(x^{2}+y^{2}\right)\cos\left(\nu t\right)-2xy}{\sqrt{2}\,w_{N}^{2}\sin\left(\nu t\right)}\right] \nn\\
%\eea
\beq
\label{eq:propagatorIP}
G\left(x,y;t\right)=\frac{\exp\left[-\frac{V_{0}t}{\hbar}-\frac{\left(x^{2}+y^{2}\right)\cos\left(\nu t\right)-2xy}{\sqrt{2}\,w_{N}^{2}\sin\left(\nu t\right)}\right]}{\left[\sqrt{2}\,\pi\sin\left(\nu t\right)\right]^{1/2}w_{N}} \;,
\eeq
where $w_N$ is the characteristic scale given in Eq. \eqref{eq:wNdef}.

We now obtain the kernel in the critical region $x \sim w_N$, $y \sim w_N$ and $\mu_1 = O\left(1\right)$, see the central square in Fig. \ref{fig:diagram}. Substituting Eq.~(\ref{eq:propagatorIP}) into~(\ref{laplace_inverse}), using the definition of $\mu_1$ in \eqref{eq:mu1def} and rescaling the time variable
we find that the kernel takes the scaling form
\beq
\label{eq:K_scaling}
K_{\mu}\left(x,y\right) \simeq \frac{\mathcal{K}^{IP}\left[\frac{x}{w_{N}},\frac{y}{w_{N}} ; \sqrt{\frac{m}{2V_{2}}}\frac{\left(\mu-V_{0}\right)}{\hbar}\right]}{w_{N}},
\eeq
where the scaling function $\mathcal{K}^{IP}(X,Y;\mu_1)$ (IP standing for inverted parabola) is a universal family of kernels, parametrized by $\mu_1$, given by
\bea
\label{eq:K_IP_def}
&&\mathcal{K}^{IP}\left(X,Y;\mu_{1}\right)=\int_{\Gamma}\frac{d\tau}{2^{5/4}\pi^{3/2}i\tau\sqrt{\sin\tau}} \nn\\
&&\times \exp\left[\mu_{1}\tau+\frac{\left(X+Y\right)^{2}\tan\left(\frac{\tau}{2}\right)-\left(X-Y\right)^{2}\cot\left(\frac{\tau}{2}\right)}{2\sqrt{2}}\right]. \nn\\
\eea
The contour in Eq.~(\ref{eq:K_IP_def}) and in all other Bromwich integrals below is
\beq
\label{eq:Bromwich_contour}
\Gamma : \quad \tau=\tau_{0}+is \quad \text{where} \quad \tau_{0}\in\left(0,\pi\right)
\eeq
is arbitrary and $s$ runs from minus infinity to plus infinity, see Appendix \ref{appendix:Bromwich} for a more detailed analysis. The rescaled kernel $\mathcal{K}^{IP}\left(X,Y;\mu_{1}\right)$ is plotted as a function of $X$ for fixed $Y$ ($Y=0$ and $Y=2$) and for $\mu_1=0$ in Fig.~\ref{fig:kernel0}. This plot,  and the other plots below, were made using a numerical evaluation of the integrals~(\ref{eq:K_IP_def}) and ~(\ref{eq:F_IP_def}) along the contour~(\ref{eq:Bromwich_contour}) with $\tau_0 = \pi/2$, but we checked that the result is independent of $\tau_0 \in (0, \pi)$. The above family of IP kernels is characteristic of double well potentials near the critical point and describes a new universality class of quantum correlations, different from that of 
the sine and Airy kernels.

In particular, from Eq.~(\ref{coinciding}) at coinciding points $x=y$ we immediately obtain that the local average density takes the scaling form near $x=0$  
\bea
\label{eq:rho_scaling}
\antiquad N\rho_{N}\left(x\right)&=&K_{\mu}\left(x,x\right) \simeq \frac{F_{1}^{IP}\left[\frac{x}{w_{N}};\frac{\sqrt{m}\,\left(\mu-V_{0}\right)}{\hbar\sqrt{2V_{2}}}\right]}{w_{N}} \;,
\eea
where the family of scaling functions is given by
\bea
\label{eq:F_IP_def}
\antiquad F_{1}^{IP}\left(X;\mu_{1}\right)=\int_{\Gamma}\frac{\exp\left[\mu_{1}\tau+\sqrt{2}\,X^{2}\tan\left(\frac{\tau}{2}\right)\right]d\tau}{2^{5/4}\pi^{3/2}i\tau\sqrt{\sin\tau}}\,.
\eea
The function $F_{1}^{IP}\left(X;\mu_{1}\right)$
is plotted in Fig.~\ref{fig:density} (a) for different values of $\mu_1$.

%
%Since the expression~(\ref{eq:propagatorIP}) is only valid at $0<t<t_1 \equiv \pi\sqrt{m/\left(2V_{2}\right)}$, the contour in Eq.~(\ref{laplace_inverse}) must be taken to be $t=t_{0}+is$ where $s$ runs from minus infinity to infinity and $t_0$ is chosen arbitrarily within the interval $t_{0}\in\left(0,t_{1}\right)$.
%Accordingly, the contour in Eqs.~(\ref{eq:K_IP_def}) and~(\ref{eq:F_IP_def}) must be $\tau=\tau_{0}+is$ where $s$ runs from minus infinity to infinity and $\tau_{0}\in\left(0,\pi\right)$.
%

We now discuss the various asymptotics of this formula by referring to the Fig.~\ref{fig:diagram}
where the different regimes of correlations are illustrated. For $\mu_1 + X^2/\sqrt{2} \gg 1$,
i.e. the region denoted "sine kernel" in Fig.~\ref{fig:diagram}, the integral in \eqref{eq:F_IP_def} is dominated
by small $\tau$ (taking $\tau_0=0^+$) giving
$F_{1}^{IP}(X,\mu_{1}) \simeq 
F_{1}^{\rm bulk}(X;\mu_{1})$ with 
(see Appendix \ref{appendix:bulk_limit} for details)
\bea
\label{eq:F_IP_bulk_approx2}
\antiquad F_{1}^{{\rm bulk}}\left(X;\mu_{1}\right)=\frac{2^{3/4}\sqrt{\mu_{1}+\frac{X^{2}}{\sqrt{2}}}}{\pi}\Theta\left(\mu_{1}+\frac{X^{2}}{\sqrt{2}}\right), \nn\\
\mu_{1}+\frac{X^{2}}{\sqrt{2}}\gg1,
\eea
which, inserted into \eqref{eq:rho_scaling} coincides with the bulk density result \eqref{bd}.
This function is plotted in Fig.~\ref{fig:density} (b).
%(TODO: exchange between Figs. \ref{fig:density} (a) and (b)?)
In particular, for $\mu_1=0$ one obtains the large $\left|X\right|$ behavior 
\beq
\label{eq:density_large_X}
\!\!F_{1}^{IP}\left(X;\mu_{1} = 0\right)\simeq\frac{2^{1/2}\left|X\right|}{\pi},\qquad\left|X\right|\gg1,
\eeq
which matches the linear behavior of the bulk density at $\mu=V_0$. 
Similarly, for $X$ and $Y$ inside the bulk, the kernel \eqref{eq:K_IP_def} converges to the sine kernel, see Appendix \ref{appendix:bulk_limit}.

Another limiting behavior is the crossover from IP to Airy indicated in blue in the
Fig.~\ref{fig:diagram}, obtained for $\mu_1 \to -\infty$ and $X$ close to one of the two edges $X = \pm X_e$ with $X_e=2^{1/4} \sqrt{|\mu_1|}$, with a distance $|X|-X_e \sim |\mu_1|^{-1/6}$.
In that limit (for the edge at $+X_e$) one has (see Appendix \ref{appendix:Airy})
%\be
%A |\mu_1|^{-1/6} F_1^{\rm IP}(2^{1/4} \sqrt{|\mu_1|} + A |\mu_1|^{-1/6} z, \mu_1) \to F_1(z) 
%\ee 
%{\red Naftali can you put the value of $A$ somewhere?} 
\beq
\label{eq:F_IP_Airy_limit}
\lim_{\mu_{1}\to-\infty}\!\frac{F_{1}^{IP}\left(2^{1/4}\left|\mu_{1}\right|^{1/2}-2^{-3/4}\left|\mu_{1}\right|^{-1/6}z;\mu_{1}\right)}{2^{3/4}\left|\mu_{1}\right|^{1/6}}\!=\!F_{1}\!\left(z\right).
\eeq
where $F_1(z)={\rm Ai}'(z)^2- z {\rm Ai}(z)^2$ is the scaling function for the 
average density at a soft edge, see \eqref{softdens}.
Similarly, at $\mu_1 \to -\infty$ and $X$ and $Y$ near the soft edge, the kernel \eqref{eq:K_IP_def} converges to the Airy kernel, see Appendix \ref{appendix:Airy}.

%{\red Q: another crossover inside the empty region. } 

\begin{figure}[ht]
	\includegraphics[width = 0.98 \linewidth]{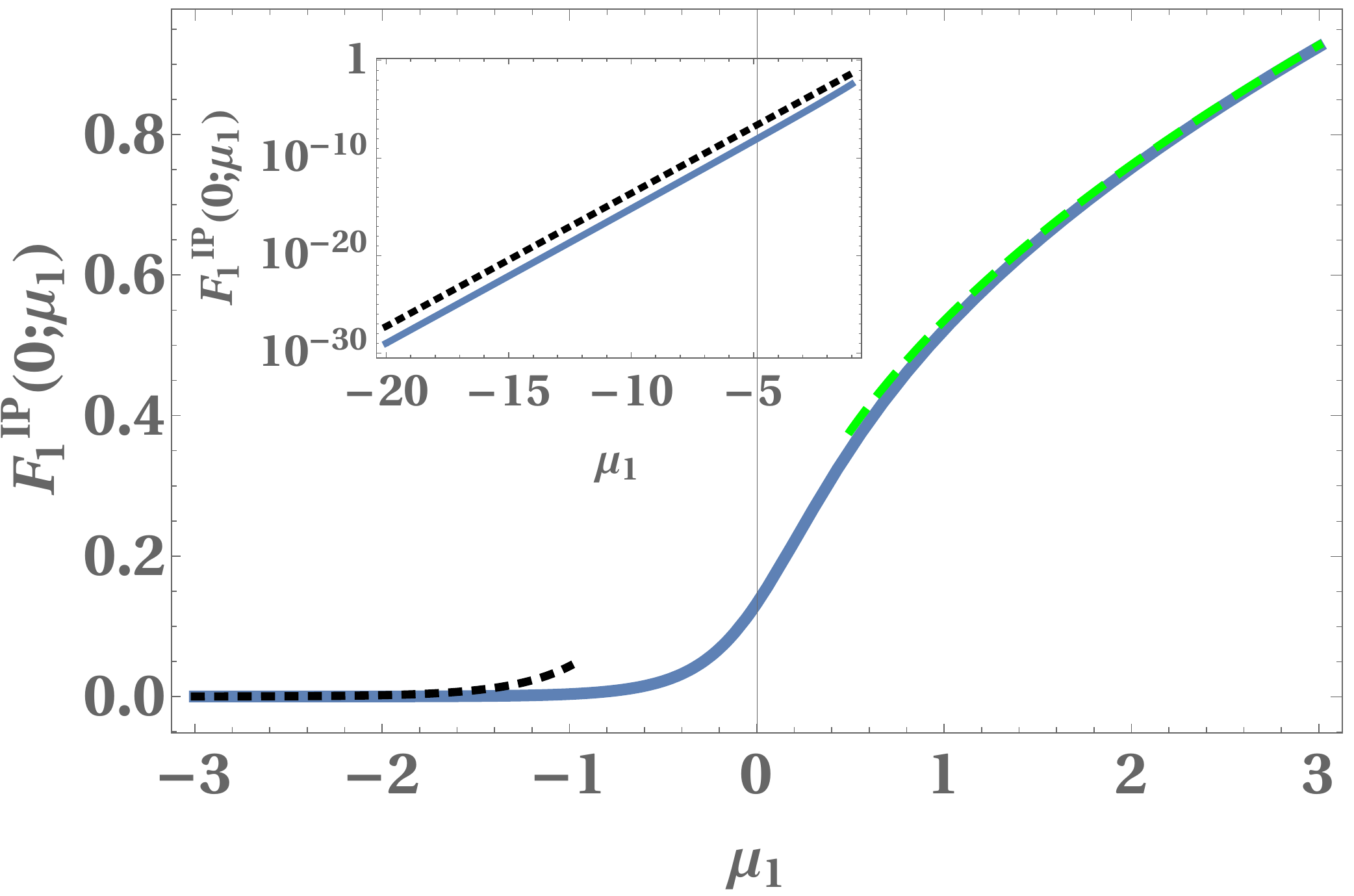}
	\caption{
		The rescaled density at the origin \eqref{eq:F_IP_at_origin} (solid line) as a function of $\mu_{1}$ together with its asymptotics~(\ref{eq:F_IP_at_origin_bulk}) and~(\ref{eq:F_IP_at_origin_WKB}) (dashed and dotted respectively). Inset: the tail $\mu_{1}\to-\infty$ plotted on a logarithmic scale. The dotted line corresponds to the asymptotic behavior given in \eqref{eq:F_IP_at_origin_WKB}} \label{fig:density_at_origin}
\end{figure}

It is interesting to note that the exact average number density never strictly vanishes at $x=0$.
In the critical region $\mu_1 = O\left(1\right)$ it is of order 
$N \rho_N(0) \sim 1/w_N$ which is of order $O\left(1\right)$. This is in contrast to the standard
soft edge where $N \rho_{N}\left(x_{e}\right) \sim 1/w_N^s$ which is of order $O(N^{1/6})$. 
The amplitude is $F_{1}^{IP}\left(X = 0;\mu_{1}\right)$
which depends continuously on $\mu_1$
\beq
\label{eq:F_IP_at_origin}
F_{1}^{IP}\left(0;\mu_{1}\right)=\int_{\Gamma}\frac{\exp\left(\mu_{1}\tau\right)d\tau}{2^{5/4}\pi^{3/2}i\tau\sqrt{\sin\tau}}\;.
\eeq
Its asymptotic behaviors are given by (see Appendix \ref{appendix:asymptotics_at_origin})
\bea
\label{eq:F_IP_at_origin_bulk}
F_{1}^{IP}\left(0;\mu_{1}\right)&\simeq&\frac{2^{3/4}\sqrt{\mu_{1}}}{\pi},\quad\mu_{1}\gg1,\\
\label{eq:F_IP_at_origin_WKB}
F_{1}^{IP}\left(0;\mu_{1}\right)&\sim& e^{\pi\mu_{1}},\qquad\; -\mu_{1}\gg1.
\eea
For the critical case $\mu_1 = 0$, $F_{1}^{IP}\left(0;0\right) = 0.13277420\dots$.
$F_{1}^{IP}\left(0;\mu_{1}\right)$ is plotted together with its asymptotics~(\ref{eq:F_IP_at_origin_bulk}) and~(\ref{eq:F_IP_at_origin_WKB}) in Fig.~\ref{fig:density_at_origin}.

\begin{figure}[ht]
	\includegraphics[width = 0.98 \linewidth]{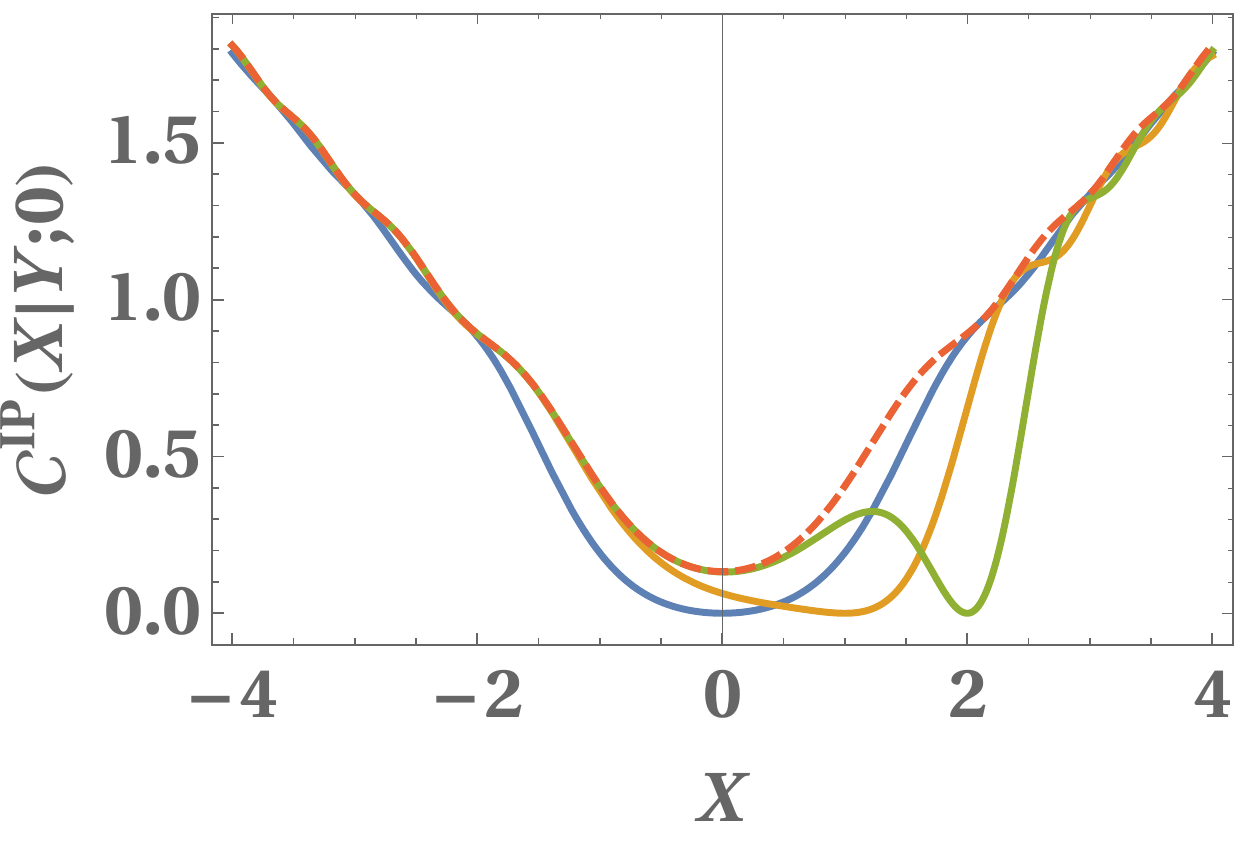}
	\caption{
		Solid: The rescaled conditional density~(\ref{eq:conditional_density}) at $X$ given that there is a particle at $Y$, for $Y=0,1,2$ and $\mu_1=0$. Dashed: unconditioned rescaled density $F_{1}^{IP}\left(X;0\right)$.} \label{fig:kernel}
\end{figure}

Another interesting physical observable, related to the kernel, is the density at $x$ conditioned on the presence of a particle at $y$, denoted by $\rho_N \left(x|y\right)$, and defined as
\be
\rho_N \left(x|y\right) = \frac{R_2\left(x,y\right)}{N (N-1) \rho_N(y)}
\ee 
which integrates to unity over $x$, i.e. $\int dx \rho_N \left(x|y\right)=1$. 
Using the determinantal expression for $R_2$ in Eq.~(\ref{determinantal_structure}), we find that 
\be
\left(N-1\right)\rho_{N}\left(x|y\right)=\frac{K_{\mu}\left(x,x\right)K_{\mu}\left(y,y\right)-\left[K_{\mu}\left(x,y\right)\right]^{2}}{K_{\mu}\left(y,y\right)} .
\ee
Near the critical point $x=0$, using the scaling properties of the kernel in Eqs. \eqref{eq:K_scaling}
and \eqref{eq:rho_scaling} we obtain for large $N$
\bea
N\rho_N \left(x|y\right)  &\simeq & \frac{\mathcal{C}^{IP}\left[\frac{x}{w_{N}}|\frac{y}{w_{N}};\frac{\sqrt{m}\,\left(\mu-V_{0}\right)}{\hbar\sqrt{2V_{2}}}\right]}{w_{N}}
\eea
where
\beq
\label{eq:conditional_density}
\mathcal{C}^{IP}\left(X|Y;\mu_{1}\right)\equiv F_{1}^{IP}\left(X;\mu_{1}\right)-\frac{\mathcal{K}^{IP}\left(X,Y;\mu_{1}\right)^{2}}{F_{1}^{IP}\left(Y;\mu_{1}\right)}. 
\eeq
The function $\mathcal{C}^{IP}\left(X|Y;\mu_{1}\right)$ is plotted in Fig.~\ref{fig:kernel} as a function of $X$ for a fixed $Y$, for the particular case $\mu_1 = 0$.
As is seen in the figure, the particle at $Y$ creates a localized ``cavity'' around $X \simeq Y$ for this conditional density. This is a manifestation of the Pauli exclusion principle.

\subsection{Representation in terms of eigenfunctions}

In this section, we use a second method to obtain the kernel and the density. We start from the expression \eqref{Kmu} of the kernel as a sum over eigenfunctions of the inverted parabola potential 
$V(x)= V_0 - V_2 x^2$. This potential does not have any bound states and hence its spectrum is continuous. Consequently, the discrete sum over the energy levels $\epsilon_k$ in the definition of the kernel in Eq. (\ref{Kmu}) needs to be replaced by an integral over $\epsilon_k$. It turns out to be convenient to express this integral in terms of the following variable $a$ defined as
\begin{eqnarray}\label{def_a_text}
a = - \frac{m^{1/2}(\epsilon_k-V_0)}{2^{1/2} \hbar V_2^{1/2}} \;.
\end{eqnarray}
Thus the integral over $\epsilon_k$ transforms into the integral over $a$, 
we obtain (see Appendix \ref{appendix:eigenfunctions} for details)
\bea
\label{eq:K_IP_eigenfunctions}
&&\antiquad \mathcal{K}^{IP}\left(X,Y;\mu_{1}\right)=2^{3/4} \!\! \sum_{\sigma\in\left\{ +,-\right\} }\nn\\
&&\qquad\qquad \int_{-\mu_{1}}^{\Lambda}\tilde{\phi}_{a,\sigma}\left(2^{3/4}X\right)\tilde{\phi}_{a,\sigma}\left(2^{3/4}Y\right)da,\label{kipd}
\eea
where the real eigenfunctions $\phi_{a,\pm}$ are labelled by $a$ and $\sigma = \pm$ -- the later is a parity index that we need to label
the eigenfunction (see Appendix \ref{appendix:eigenfunctions}). In Eq. (\ref{eq:K_IP_eigenfunctions}), the 
upper cut-off 
$$\Lambda=-\frac{m^{1/2}\left(\epsilon_{\min}-V_{0}\right)}{2^{1/2}\hbar V_{2}^{1/2}}$$
corresponds to the maximal value of $a$ in (\ref{def_a_text}) -- this is associated to $\epsilon_k = \epsilon_{\min}$ corresponding to the single particle ground state energy. Note that in the expression for $\Lambda$, $V_2$ and $\epsilon_{\min}$ are both of order $O(1)$ while $V_0 = O(\mu)$ near the critical point. Thus in the large $\mu$ limit, $\Lambda \to \infty$ and we can extend the upper limit of the integral in Eq. (\ref{eq:K_IP_eigenfunctions}) to $+\infty$ provided the integral converges (which is indeed the case here). It turns out that the eigenfunctions can be explicitly computed as follows \citep{Barton1986}
\bea
\label{eq:phiplus}
\tilde{\phi}_{a,+} \! \left(x\right) \! &=& \! \frac{e^{-\frac{\pi a}{4}} \! \left|\Gamma\left(\frac{1}{4}+\frac{ia}{2}\right)\right|}{2^{5/4}\pi}e^{-\frac{ix^{2}}{4}}\!\,_{1}F_{1} \! \left(\frac{1}{4}-\frac{ia}{2},\frac{1}{2},\frac{ix^{2}}{2}\right),\nn\\\\
\label{eq:phiminus}
\tilde{\phi}_{a,-} \! \left(x\right) \! &=& \! \frac{e^{-\frac{\pi a}{4}} \! \left|\Gamma\left(\frac{3}{4}+\frac{ia}{2}\right)\right|}{2^{3/4}\pi}xe^{-\frac{ix^{2}}{4}}\!\,_{1}F_{1} \! \left(\frac{3}{4}-\frac{ia}{2},\frac{3}{2},\frac{i x^{2}}{2}\right) , \nn\\
\eea
where $_{1}F_{1}$ is the Kummer hypergeometric function \citep{wolfram_hypergeometric}.
In particular, the rescaled density is given by
\bea
\label{eq:F_IP_eigenfunctions}
&&\antiquad \!\!\!  F_{1}^{IP}\left(X;\mu_{1}\right)=\mathcal{K}^{IP}\left(X,X;\mu_{1}\right)\nn\\
&&\qquad\quad\;\;=2^{3/4}\!\!\!\! \sum_{\sigma\in\left\{ +,-\right\} }\int_{-\mu_{1}}^{\infty}\left[\tilde{\phi}_{a,\sigma}\left(2^{3/4}X\right)\right]^{2} \! da\;.
\eea
We verified numerically that the formulae (\ref{eq:K_IP_eigenfunctions}) and (\ref{eq:F_IP_eigenfunctions}) coincide with (\ref{eq:K_IP_def}) and (\ref{eq:F_IP_def}) respectively. Futhermore, in the form given in  Eq. (\ref{eq:K_IP_eigenfunctions}), one may verify directly that the kernel is reproducible in the sense defined in Eq. (\ref{reprod}).
Showing analytically that the formulae coincide seems quite challenging. 
%This coincidence appears to be equivalent to a Nicholson-type integral \citep{Nicholson_integral}.
The above results can also be derived via a Green's function method which we explain in Appendix \ref{Green's}.

\section{Counting statistics in the critical regime}
\label{sec:number_variance}

In this section we briefly discuss some applications of this new IP kernel to describe the
fluctuations of the number of fermions in a given interval within the critical regime 
of the double well potential. 

\begin{figure*}[ht]
	\includegraphics[width = 0.48 \linewidth]{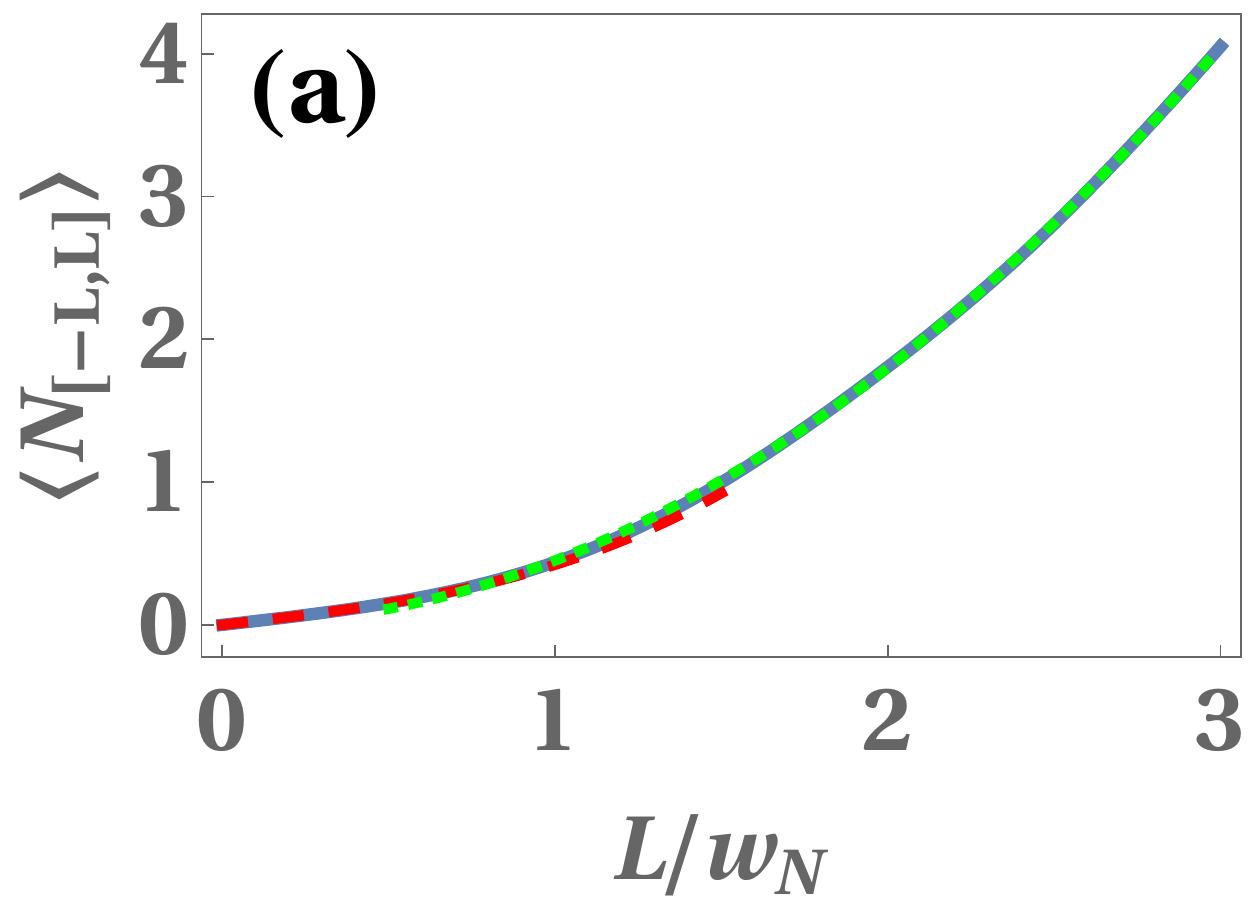}
	\includegraphics[width = 0.50 \linewidth]{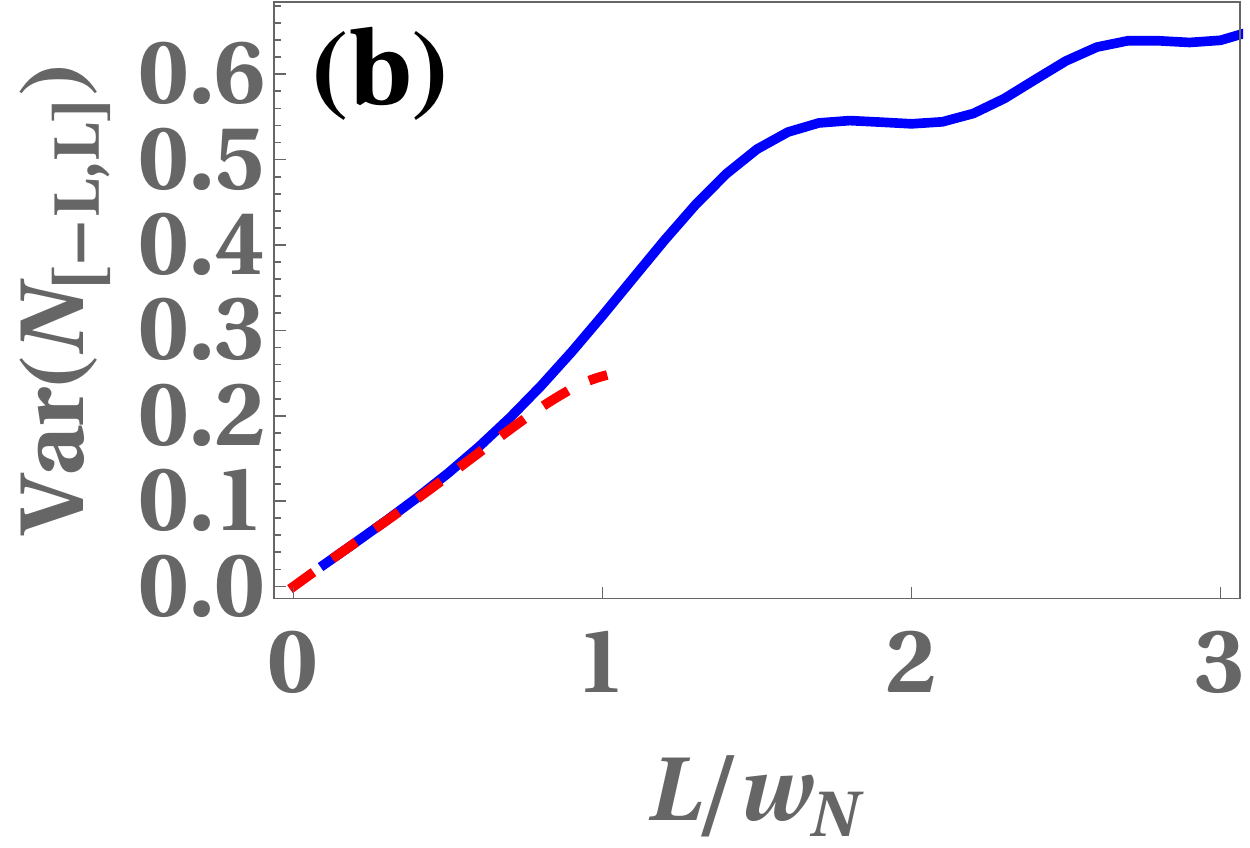}
	\caption{
		(a) Solid line: mean number of fermions in the interval $\left\langle N_{\left[-L,L\right]}\right\rangle$ as a function of the rescaled length $\ell=L/w_N$ {for $\mu_1 = 0$}, see Eq.~(\ref{eq:mean_scaling}).
		{Dashed and dotted lines: small- and large-$\ell$ approximations, Eqs.~\eqref{eq:mean_small_L} and \eqref{eq:mean_large_L} respectively.}
		(b) Solid line: Numerical evaluation of the variance of the number of fermions in the interval $\text{Var}\left(N_{\left[-L,L\right]}\right)$ as a function of the rescaled length {for $\mu_1 = 0$}, see Eq.~(\ref{eq:variance_scaling}). {Dashed} line: The small-$\ell$ asymptotic of the number variance, describing the variance of a Bernoulli random variable.
	} \label{fig:numberVariance}
\end{figure*}

We start with evaluating the average number of fermions within an interval $\mathcal{I} \subset \mathbb{R}$
which is given by $\left\langle N_{\mathcal{I}}\right\rangle =\int_{\mathcal{I}}N\rho_{N}\left(x\right)dx$. 
Choosing ${\cal I} = [-L,L]$, and using the formula \eqref{eq:rho_scaling} and \eqref{eq:F_IP_def} for the average density, it takes the scaling form
\beq
\label{eq:mean_scaling}
\left\langle N_{\left[-L,L\right]}\right\rangle \simeq\mathcal{N}\left[\frac{L}{w_{N}}\,;\frac{\sqrt{m}\,\left(\mu-V_{0}\right)}{\hbar\sqrt{2V_{2}}}\right],\quad L=O\left(w_{N}\right)
\eeq
where the scaling function $\mathcal{N}\left(\ell;\mu_{1}\right)$ is given by
\bea
\label{eq:number_mean}
\!\! \mathcal{N}\left(\ell;\mu_{1}\right)& \! =&  \int_{-\ell}^{\ell}F_{1}^{IP}\left(X;\mu_{1}\right)dX \nn\\
\! =&&\!\!\!\!\! \!\int_{\Gamma} \! \frac{e^{\mu_{1}\tau}d\tau}{2^{5/4}\pi^{3/2}i\tau\sqrt{\sin\tau}}\int_{-\ell}^{\ell}\!\!\exp\left[\sqrt{2}\,X^{2}\tan\left(\frac{\tau}{2}\right)\right] \! dX \nn\\
&\!=& \int_{\Gamma}\frac{\text{erfi}\left[2^{1/4}\sqrt{\tan\left(\frac{\tau}{2}\right)}\,\ell\right]e^{\mu_{1}\tau}d\tau}{2^{3/2}\pi i\tau\sqrt{\sin\tau\tan\left(\frac{\tau}{2}\right)}}
\eea
where $\text{erfi}\left(z\right)=\left(2/\!\sqrt{\pi}\right)\int_{0}^{z}e^{t^{2}}dt$ is the imaginary error function.
In the small $\ell \ll 1$ limit one obtains, via a Taylor expansion, the asymptotics
\beq
\label{eq:mean_small_L}
\mathcal{N}\left(\ell\ll1;\mu_{1}\right)\simeq2F_{1}^{IP}\left(0;\mu_{1}\right)\ell+
A \ell^{3} + O(\ell^5)
\eeq
where
\beq
A=\frac{\left.\partial_{X}^{2}F_{1}^{IP}\left(0;\mu_{1}\right)\right|_{X=0}}{3}=\int_{\Gamma}\frac{2^{1/4}\tan\left(\frac{\tau}{2}\right)\exp\left(\mu_{1}\tau\right)d\tau}{3 \pi^{3/2}i\tau\sqrt{\sin\tau}}.
\eeq
At large $\ell$, but within the critical region, 
%\sout{it behaves as $\mathcal{N}\left(\ell\ll1,\mu_{1}\right) \simeq \frac{2^{3/4}}{\sqrt{\pi}} \ell^2$.}
{we obtain, by plugging Eq.~\eqref{eq:density_large_X} into \eqref{eq:number_mean},}
\beq
\label{eq:mean_large_L}
\mathcal{N}\left(\ell\gg1;\mu_{1}\right)\simeq\frac{2^{1/2}\ell^{2}}{\pi}.
\eeq
The function $\mathcal{N}\left(\ell;\mu_{1}\right)$ is plotted
in Fig. \ref{fig:numberVariance} {(a) for $\mu_1 = 0$} and compared to its asymptotics.

An important and experimentally observable, which 
characterizes the quantum fluctuations, is the number variance $\text{Var}\left(N_{\mathcal{I}}\right) \! = \! \left\langle N_{\mathcal{I}}\right\rangle^2 \! - \left\langle N_{\mathcal{I}}^2\right\rangle$. For any determinantal point process, it is given in terms of the kernel as \citep{mehta}
\bea
%\label{eq:Vardef}
\text{Var}\left(N_{\mathcal{I}}\right)&=& \text{Tr}\left(K\right)-\text{Tr}\left(K^{2}\right) \nn \\
&=& \left\langle N_{\mathcal{I}}\right\rangle - \int_{\mathcal{I}}\int_{\mathcal{I}}\left[K\left(x,y\right)\right]^{2}dxdy \, .
\eea
Using Eq.~(\ref{eq:K_scaling}) we obtain that the number variance takes the scaling form 
\beq
\label{eq:variance_scaling}
\text{Var}\left(N_{\left[-L,L\right]}\right)\simeq\mathcal{V}\left[\frac{L}{w_{N}};\frac{\sqrt{m}\,\left(\mu-V_{0}\right)}{\hbar\sqrt{2V_{2}}}\right],\quad L=O\left(w_{N}\right),
\eeq
where the scaling function can be expressed in terms of the inverted parabola kernel
as
\beq
\label{eq:scaled_number_variance}
\mathcal{V}\left(\ell;\mu_{1}\right)=\mathcal{N}\left(\ell;\mu_{1}\right)-\int_{-\ell}^{\ell}\int_{-\ell}^{\ell}\left[\mathcal{K}^{IP}\left(X,Y;\mu_{1}\right)\right]^{2}\!dXdY.
\eeq
We have evaluated numerically this formula, which is plotted
in Fig.~\ref{fig:numberVariance} {(b) for $\mu_1 = 0$}. {At small rescaled length $\ell= L/w_N$,
there can be at most one fermion in the interval. Consequently the number of fermions in $[-L,L]$ becomes
a Bernoulli random variable,  and hence }
$\text{Var}\left(N_{\left[-L,L\right]}\right)\simeq\left\langle N_{\left[-L,L\right]}\right\rangle -\left\langle N_{\left[-L,L\right]}\right\rangle ^{2}$ , an approximation which is verified in the plot in Fig. \ref{fig:numberVariance} {(b)}.

Finally, another interesting observable is the ``hole'' probability -- the probability that no fermions are observed within the interval -- which is given by the Fredholm determinant \citep{J05,Bo11}
\beq
\mathbb{P}\left(N_{\mathcal{I}}=0\right)=\text{Det}\left(I-P_{\mathcal{I}}K_\mu P_{\mathcal{I}}\right)
\eeq
where $P_{\mathcal{I}}$ is the projector on $\mathcal{I}$, such that $P_{\mathcal{I}}(x)=1$ if 
$x\in\ensuremath{\mathcal{I}}$ and $P_{\mathcal{I}}(x)= 0$ if $x\notin{\mathcal{I}}$. 
We leave the evaluation of this Fredholm determinant for future investigations. We recall that in the case
of the standard soft edge, the corresponding hole probability (computed from the Airy kernel)
can be expressed in terms of the solution of a Painlev{\'e} II equation. It would be interesting to
investigate whether the hole probability for the IP kernel {also satisfies some similar kind of  differential equation}.

\section{Discussion and conclusion}
\label{sec:discussion}

\subsection{Range of observability of the inverted parabola kernel}
\label{sec:NasFunctionOfMu}

It would be very interesting to observe the predictions of this paper in experiments on cold atoms.
For a potential of the form considered here $V(x)= V_0 v\left(x/x_{0}\right)$, one needs first to tune the
chemical potential to $\mu \approx V_0$. An important issue is whether the critical regime 
$\mu_1 = O\left(1\right)$ where $\mu_1$ is defined in Eq. \eqref{eq:mu1def} is observable without the need to further tune the parameters of the potential, i.e. the function $v(z)$. In this subsection we show that in the semi-classical limit this is indeed possible, by choosing the number of particles $N$ appropriately. 
%That is, it is not necessary to fine tune the parameters of the potential itself.

To this aim we need to estimate how the number of fermions $N$ in the well behaves as
a function of $\mu$ for $\mu \approx V_0$. Let us denote $N_0$ the number 
of fermions exactly at $\mu=V_0$.
It can be approximated from the bulk density, by inserting the form in Eq.~(\ref{eq:potentialScaling}) for $V(x)$
in the formula~(\ref{eq:NasFunctionOfMu}), which gives
\beq
\label{eq:N0def}
N_{0}\equiv\left.N\right|_{\mu=V_{0}}\simeq\frac{C\left(2mV_{0}\right)^{1/2}x_{0}}{\pi\hbar}
\eeq
where
\beq
C=\int_{-\infty}^{\infty}dz\left[1-v\left(z\right)\right]^{1/2}\Theta\left[1-v\left(z\right)\right]
\eeq
is a dimensionless constant, of order unity, that depends on the global properties of the potential.

We can now approximate the number of fermions at chemical potential $\mu \neq V_0$ as
\be \label{eq00} 
N=N_{0}+\left(\mu-V_{0}\right)\left.\frac{dN}{d\mu}\right|_{\mu=V_{0}} \;.
\ee
To estimate $\left.\left(dN/d\mu\right)\right|_{\mu=V_{0}}$, we take the derivative of Eq.~(\ref{eq:NasFunctionOfMu}) with respect to $\mu$. This yields
%{\blue Naftali: I changed $\simeq$ to $=$ in the second line of this equation:}
\bea
%\label{eq:dNdmu}
 \frac{dN}{d\mu}&\simeq&\frac{\sqrt{m}}{\pi\hbar}\int_{-\infty}^{\infty}\frac{1}{\sqrt{2\left[\mu-V\left(x\right)\right]}_+} \, dx \nn \\
 &=& \frac{x_{0}\sqrt{m}}{\sqrt{2V_{0}}\pi\hbar}\int_{-\infty}^{\infty}\frac{1}{\sqrt{\left[1-v(z)+\frac{\mu-V_{0}}{V_{0}}\right]_{+}}}\,dz \;,
\eea
where we used that $V\left(x\right)=V_{0}v\left(x/x_{0}\right)$, where $1/\!\sqrt{x_+}= \Theta(x)/\sqrt{x}$. At $\mu_1=0$ this integral is dominated by the region $z \ll 1$, where one can use the approximation $1-v(z) \simeq - v''(0) z^2/2$.
The logarithmic divergence is cutoff at a scale $z=w_N/x_0$ where one enters the critical regime
where the density remains nonzero. One obtains 
%\bea
%\frac{dN}{d\mu} \simeq\frac{\sqrt{2m}}{\pi\hbar\sqrt{V_{2}}}\left[\ln\left(z+\sqrt{z^{2}+\alpha\mu_{1}}\right)\right]_{w_{N}/x_{0}}^{z_{c}} \;,
%\eea
\beq
\!\! \frac{dN}{d\mu} \! \simeq \! \frac{\sqrt{2m}}{\pi\hbar\sqrt{V_{2}}}\left[\ln\left(\! z+\sqrt{z^{2}+\frac{\sqrt{2}\,w_{N}^{2}\mu_{1}}{x_{0}^{2}}}\right)\right]_{w_{N}/x_{0}}^{z_{c}},
\eeq
where $z_c$ is a cutoff of order unity. 
For $\mu_1=O\left(1\right)$ and $w_N/x_0 \ll 1$ this gives the
following leading order estimate (up to {an additive} term of order unity)
\beq
\label{eq:dNdmu2}
\frac{dN}{d\mu} \simeq\frac{\sqrt{2m}}{\pi\hbar\sqrt{V_{2}}}\int_{w_{N}}^{x_{0}}\frac{dx}{x}=\frac{\sqrt{2m}}{\pi\hbar\sqrt{V_{2}}}\ln\left(\frac{x_{0}}{w_{N}}\right) \;.
\eeq
Using \eqref{eq00} and the fact that $x_0/w_N \! \sim \! \sqrt{N_0}$ (e.g. from \eqref{eq:N0def}
and \eqref{eq:V2def} and consistent with Eq. \eqref{scale}) 
together with the definition \eqref{eq:mu1def} of $\mu_1$ 
we finally obtain the connection between $N$ and $\mu$ at $N \simeq N_0$ 
\beq
\label{eq:N_of_mu1}
N\left(\mu\right)\simeq N_{0}+\frac{\ln N_{0}}{\pi}\mu_{1} \;.
\eeq
Hence if we increase the number of fermions by one, in the critical regime the parameter $\mu_1$ 
increases by about $1/\ln N$ which is small. Hence the critical regime is stable by adding a few particles
and can be probed continuously at large $N$. 
%%
%%Finally, the energy levels of the Hamiltonian~(\ref{ham}), at $\epsilon_k \simeq V_0$, are obtained from Eqs.~(\ref{eq:mu1def}) and~(\ref{eq:N_of_mu1}) by replacing $N$ by $k$ and $\mu$ by $\epsilon_k$, implying that $\epsilon_k$ can be treated as continuous in the limit $N_0 \gg 1$. This will be useful to us in the calculation of the kernel using the eigenfunctions, see Eq.~(\ref{eq:K_IP_eigenfunctions}) and Appendix \ref{appendix:eigenfunctions} below.

\subsection{Extensions to finite temperature and higher dimensions}

The calculations presented in this paper can be extended to finite temperature
in the grand canonical ensemble.
The simplest way to obtain the finite temperature kernel from the zero
temperature kernel $K_\mu\left(x,y\right)$ computed here is to use the formula (240)
in \cite{fermions_review}, which reads (for the particular case $d=1$)
\beq
\label{eq:nonzero_temperature_K_def}
K_{\tilde{\mu}}\left(x,y\right)=\int_{0}^{\infty}d\mu\frac{\partial K_{\mu}\left(x,y\right)}{\partial\mu}\frac{1}{1+e^{\beta\left(\mu-\tilde{\mu}\right)}},
\eeq
where $\beta = 1/T$ is the inverse temperature 
and $\tilde{\mu}$ is the chemical potential, related to $N$ and $T$ through
\beq
N=\sum_{k}\frac{1}{1+e^{\beta\left(\epsilon_{k}-\tilde \mu\right)}}.
\eeq
There is
family of IP kernels indexed by a dimensionless inverse temperature (see below).
The relevant temperature scale $T_{\rm IP}$ in the critical region 
can be estimated by comparing the thermal de Broglie wavelength 
$\lambda_{T}=\hbar\sqrt{2\pi/\left(mT\right)}$ 
and the edge width $w_N$ defined in Eq.~\eqref{eq:wNdef}. 
The two length scales are of the same order 
$\lambda_{T_{\rm IP}} \simeq w_N$ for temperatures of order $T\sim T_{\rm IP}$
with 
\be
T_{\rm IP} = \frac{\hbar V_{2}^{1/2}}{m^{1/2}}  \;.
\ee
This can also be obtained by equating the potential energy scale associated to
the critical width $V_2 w_N^2$ with temperature $T$. Remarkably $ T_{{\rm IP}}=O\left(N^{0}\right)$, in contrast to the standard soft edge 
where the relevant temperature scale is $T\sim N^{1/3}$. Hence the critical regime in the double-well potential is more 
sensitive to thermal fluctuations.
Plugging the inverse-parabola kernel \eqref{eq:K_scaling} into Eq.~\eqref{eq:nonzero_temperature_K_def}, we obtain
	\beq
	K_{\tilde{\mu}}\left(x,y\right)\simeq\frac{\mathcal{\tilde{K}}^{IP}\left[\frac{x}{w_{N}},\frac{y}{w_{N}};\sqrt{\frac{m}{2V_{2}}}\frac{\left(\tilde{\mu}-V_{0}\right)}{\hbar}\right]}{w_{N}}
	\eeq
	where  
	\beq
	\mathcal{\tilde{K}}^{IP}\!\left(X,Y;\tilde{\mu}_{1}\right)\!=\!\int_{-\infty}^{\infty}\!\frac{d\mu_{1}}{1+e^{\sqrt{2}\,b\left(\mu_{1}-\tilde{\mu}_{1}\right)}}\frac{\partial\mathcal{K}^{IP}\!\left(X,Y;\mu_{1}\right)}{\partial\mu_{1}} \;, \label{KT} 
	\eeq
where $b=T_{\rm IP}/T$ is the dimensionless temperature. The scaled kernel is
thus a continuously varying function of the parameter $b$. In the limit $b \to +\infty$ 
\eqref{KT} converges to the zero temperature kernel \eqref{eq:K_IP_def}.

{Furthermore,} our calculation can be extended to higher dimensions $d>1$ for any quadratic potential
of the type 
{$V\left({\mathbf{x}}\right)=V_0 + \sum_{i=1}^{d}\epsilon_{i}\omega_{i}^{2}x_{i}^{2}$}, 
where $\epsilon_i = \pm 1$. 
The propagator for such potentials can be obtained
by analytical continuation in the frequencies $\omega_i$'s of the propagator of the $d$-dimensional harmonic oscillator, as was done in this paper
for $d=1$. Near a stationary
point of an arbitrary multidimensional potential
one can always perform a Taylor expansion up to quadratic order and
use the expression for the propagator to compute the kernel using the general
relation 
Eq.~\eqref{laplace_inverse} between the Euclidean propagator and the kernel. 
The resulting rescaled kernel will be a multidimensional generalization of the IP kernel
unveiled in this work. The case of a local maximum will be qualitatively
similar to the present case, with an (ellipse-like) edge curve disappearing as $\mu \to V_0^-$.
The case of a saddle point however does not exhibit an edge, but rather a singularity in
the bulk density at $\mu=V_0$, which will be rounded by quantum fluctuations in the
critical region.

\subsection{Conclusion}

In conclusion, we have studied a gas of non interacting spinless fermions of mass $m$ in one dimension, at zero temperature, in a double well potential $V(x)$, with a local maximum at $x=0$.
 As a function of the Fermi energy $\mu$, the bulk average density of fermions undergoes a transition from having two disjoint supports centered 
around the two minima of the potential for $\mu<V_0=V(0)$, to a single support for $\mu>V_0$. 
Exactly at the critical point $\mu=V_0$, the two supports merge at the local maximum $x=0$.
Zooming {in} close to this critical point, we have shown that the correlations between the positions of the fermions are described by a new universal scaled kernel, which we called the inverted parabola (IP) kernel,
and for which we have obtained two equivalent analytical expressions. 
Our main results for the phase diagram are summarized in the Fig.~\ref{fig:diagram}. We have shown that
the novel critical regime is characterized by two scales: an energy scale, $\mu-V_{0}\sim\hbar\sqrt{\left|V''\left(0\right)\right|}m^{-1/2}$, and a length scale $w_{N}\sim\hbar^{1/2}\left[m\left|V''\left(0\right)\right|\right]^{-1/4}$.

While the bulk density vanishes linearly around $x=0$ at the critical point $\mu=V_0$,
we have shown that the exact density at the critical point remains finite. We have calculated, in the critical
regime, the scaling function which describes this density, as well as 
the conditional two-point correlation function,
and the {mean and} variance of the number of fermions in a fixed interval around $x=0$. 
Although the effects unveiled here require a tuning of the
chemical potential $\mu$ near $V_0$, we have shown that the width of the critical region 
is sufficiently broad to be explored in experiments.

We have discussed the effect of a finite temperature $T>0$, which leads to
another, universal kernel depending on the dimensionless parameter $b=T_{\rm IP}/T$,
which we {calculated}. The thermal fluctuations modify the 
critical behavior when the temperature is raised to the characteristic temperature scale 
$T_{{\rm IP}}\sim\hbar\left|V''\left(0\right)\right|^{1/2}\!/m^{1/2}$. We have also indicated how to extend the present
results to similar critical regimes in higher dimensional potentials, near a local
maximum or a saddle point. 

Several interesting theoretical questions remain, notably whether this new IP kernel can be related to random matrix theory, and whether its associated Fredholm determinant, which determines hole probability and the full counting fermion statistics, satisfies some non linear differential equation. Finally, it would be also interesting to investigate similar questions
when two supports of the density merge,
in the case of the dynamics of non interacting fermions \cite{us_dynamics, dubail_dyn}.

\acknowledgments

NRS acknowledges support from the Yad Hanadiv fund (Rothschild fellowship).
This research was supported by ANR grant ANR-17-CE30-0027-01 RaMaTraF.

\appendix

\section{The Bromwich contour}
\label{appendix:Bromwich}
The quantum operators corresponding to the kernel~(\ref{Kmu}) and the propagator~(\ref{eq:Gdef}) are
\beq
\hat{K}_{\mu} =\sum_{{\bf k}}\Theta\left(\mu-\epsilon_{{\bf k}}\right)\ket{\psi_{{\bf k}}}\bra{\psi_{{\bf k}}}
\eeq
and
\beq
\hat{G}\left(t\right)=\sum_{{\bf k}}\exp\left(-\frac{\epsilon_{{\bf k}}t}{\hbar}\right)\ket{\psi_{{\bf k}}}\bra{\psi_{{\bf k}}}
\eeq
respectively.
We immediately see that $\hat{G}\left(t\right)$ is defined for $-\infty < t < +\infty$, and only its matrix elements in the position representation~(\ref{eq:propagatorIP}) are confined to the interval $0 < t < t_1 \equiv \pi\sqrt{m/\left(2V_{2}\right)}$.
Writing Eq.~(\ref{laplace_inverse}) in operator form, 
\begin{equation}
\label{laplace_inverse_hat_app}
\hat{K}_{\mu}=\int_{\Gamma}\frac{dt}{2\pi it}\exp\left(\frac{\mu t}{\hbar}\right)\hat{G}\left(t\right),
\end{equation}
it is clear that the only pole of the integrand in Eq.~(\ref{laplace_inverse_hat_app}) is at $t=0$, so that the Bromwich contour can be taken to be $t = t_0 + is$ with any choice of $t_0 > 0$, where $s$ runs from minus infinity to infinity.
However, in the derivation leading to Eqs.~(\ref{eq:K_IP_def}) and~(\ref{eq:F_IP_def}) we used the position representation of the propagator, Eq.~(\ref{eq:propagatorIP}). As a result in our calculations we must take $t_{0}\in\left(0,t_{1}\right)$.
Under the change of integration variable $\tau=t\sqrt{2V_{2}/m}$ that we used when deriving Eq.~(\ref{eq:K_IP_def}) the integration contour becomes (\ref{eq:Bromwich_contour}).
%Accordingly, the contour in Eqs.~(\ref{eq:K_IP_def}) and~(\ref{eq:F_IP_def}) must be $\tau=\tau_{0}+is$ where $s$ runs from minus infinity to infinity and $\tau_{0}\in\left(0,\pi\right)$.
%

\section{Asymptotics of the rescaled density}

\subsection{Density {and kernel} in the bulk}
\label{appendix:bulk_limit}

In the limit $\mu_{1}+\frac{X^{2}}{\sqrt{2}}\gg1$, the integral~(\ref{eq:F_IP_def}) is dominated by small values of $\tau$. As a result, we can approximate it by keeping only the linear term in $\tau$ in the exponent, leading to
\beq
\label{eq:F_IP_lineart_approx}
F_{1}^{IP}\left(X;\mu_{1}\right)\simeq\int_{\Gamma}\frac{\exp\left[\left(\mu_{1}+\frac{X^{2}}{\sqrt{2}}\right)\tau\right]d\tau}{2^{5/4}\pi^{3/2}i\tau^{3/2}}.
\eeq
Using the identity
\beq
\int_{\Gamma}\frac{dt}{2\pi it^{d/2+1}}\exp\left(zt\right)=\frac{z^{d/2}}{\Gamma\left(1+\frac{d}{2}\right)}\Theta\left(z\right),
\eeq
Eq.~(\ref{eq:F_IP_lineart_approx}) yields Eq.~\eqref{eq:F_IP_bulk_approx2}.
%\beq
%\label{eq:F_IP_bulk_approx}
%F_{1}^{IP} \! \left(X;\mu_{1}\right) \! \simeq \! \frac{2\sqrt{\mu_{1}+\frac{X^{2}}{\sqrt{2}}}}{\sqrt{\pi}}\Theta \! \left( \! \mu_{1}+\frac{X^{2}}{\sqrt{2}}\right), \;\, \mu_{1}+\frac{X^{2}}{\sqrt{2}}\gg1,
%\eeq

{Similarly, in the limit $\mu_{1}+X^{2} \! /\sqrt{2}\gg1$ and  at $X\simeq Y$ the kernel \eqref{eq:K_IP_def} reduces to
\beq
\label{eq:K_IP_sine_kernel_approx}
\mathcal{K}^{IP}\!\left(X,Y;\mu_{1}\right)\simeq\frac{\sin\left(2^{3/4}\sqrt{\mu_{1}+\frac{X^{2}}{\sqrt{2}}}\,\left|X-Y\right|\right)}{\pi\left|X-Y\right|}.
\eeq
When plugging Eq.~\eqref{eq:K_IP_sine_kernel_approx} into Eq.~\eqref{eq:K_scaling}, the result matches that of the sine kernel~(\ref{eq:sine_kernel_def}).
The approximation \eqref{eq:K_IP_sine_kernel_approx} is compared to the exact expression \eqref{eq:K_IP_def} in Fig.~\ref{fig:kernel0}.
}

\subsection{{Soft-edge asymptotic of the density and kernel}}
\label{appendix:Airy}

\begin{figure}[ht]
	\includegraphics[width = 0.98 \linewidth]{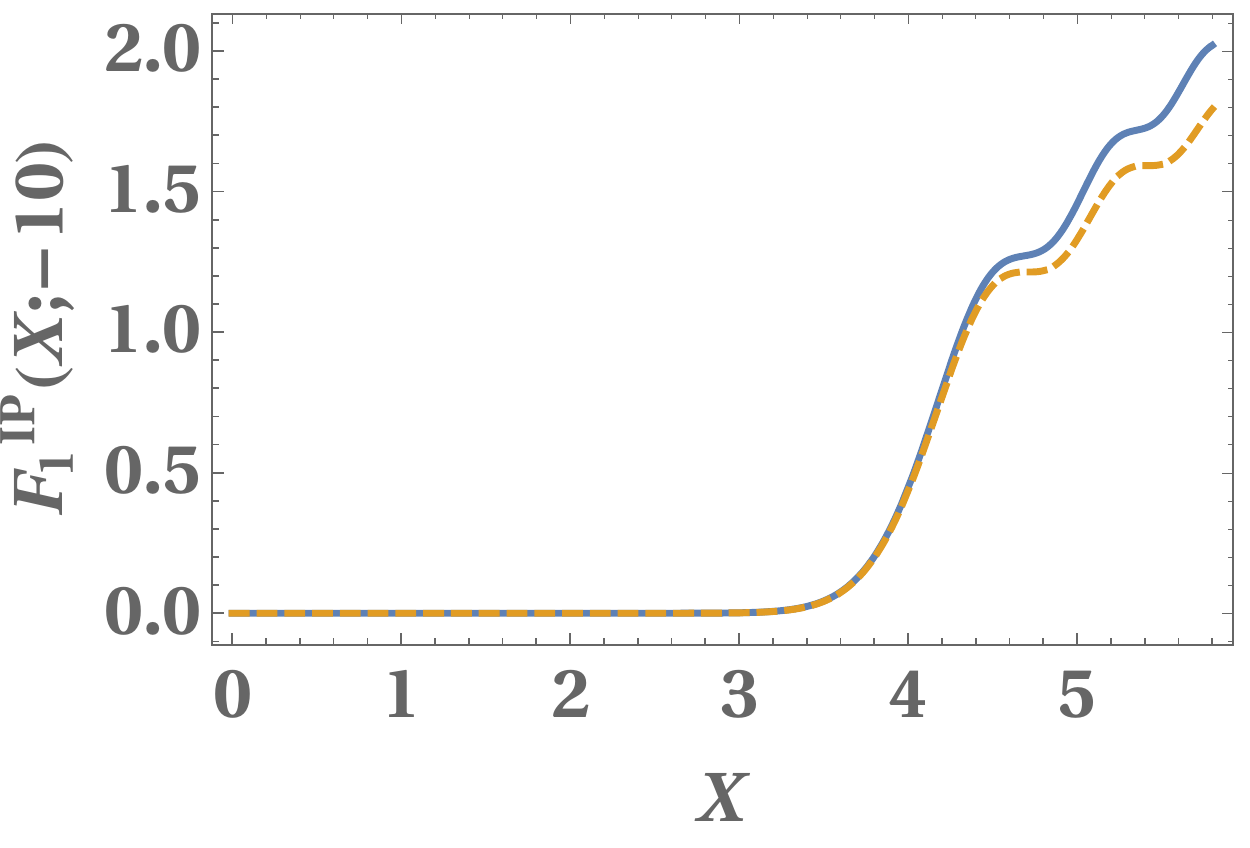}
	\caption{
		$F_{1}^{IP}\left(X;\mu_{1}=-10\right)$ (solid line) from Eq. \eqref{eq:F_IP_def}, see footnote \citep{footnote:mirror_symmetry}, together with its Airy-kernel approximation (dashed), see Eq.~\eqref{eq:F_IP_Airy_limit}.}
		\label{fig:airy_kernel}
\end{figure}

In this appendix we obtain the asymptotic behavior of the function $F_{1}\left(X;\mu_{1}\right)$ at $-\mu_{1} \gg 1$ and $X\simeq X_e = 2^{1/4}\left|\mu_{1}\right|^{1/2}$. This is the region marked ``Airy-kernel'' in Fig.~\ref{fig:diagram}.
We achieve this by expanding the term in the exponent in Eq.~\eqref{eq:F_IP_def} in leading powers of $\tau$. Denoting 
\beq
X=X_{e}+\Delta X=2^{1/4}\left|\mu_{1}\right|^{1/2}+\Delta X,
\eeq
and expanding up to order $\tau^3$, we obtain
\bea
&&\antiquad \antiquad F_{1}^{IP}\left(X;\mu_{1}\right)\simeq\int_{\Gamma}\frac{d\tau}{2^{5/4}\pi^{3/2}i\tau^{3/2}} \nn \\
&&\qquad\times\exp\left[2^{3/4}\left|\mu_{1}\right|^{1/2} \! \Delta X\tau+\frac{\tau^{2}}{12}-\mu_{1}\frac{\tau^{3}}{12}\right] \;,
\eea
where we neglected the term quadratic in $\Delta X$ because we assume that it is small, $\Delta X \ll X_e$.
Note that the term which is linear in $\tau$ vanishes at $\Delta X = 0$. This is why an expansion up to the linear order does not suffice here \citep{fermions_review}.
Rescaling $s=-\mu_{1}^{1/3}\tau$ and setting $z=-2^{3/4}\left(-\mu_{1}\right)^{1/6} \! \Delta X$ we find that the terms linear and cubic in $\tau$ are $O\left(1\right)$ but the quadratic term is $O\left(\left|\mu_{1}\right|^{-2/3}\right)$ and therefore negligible in the limit $- \mu_1 \gg 1$, so
\beq
\label{eq:F_1_IP_Airylimit_rescaled}
F_{1}^{IP}\left(X;\mu_{1}\right)\simeq\left(-\mu_{1}\right)^{1/6}\int_{\Gamma}\frac{\exp\left(-zs+\frac{s^{3}}{12}\right)ds}{2^{5/4}\pi^{3/2}is^{3/2}}.
\eeq
The integral in Eq.~\eqref{eq:F_1_IP_Airylimit_rescaled} was solved in Ref.~\cite{fermions_review}. Using its solution while taking into account the numerical factors, we finally obtain
\beq
F_{1}^{IP}\left(X;\mu_{1}\right)\simeq\left(-\mu_{1}\right)^{1/6}2^{3/4}F_{1}\left(z\right)
\eeq
leading to Eq.~\eqref{eq:F_IP_Airy_limit}.
$F_{1}^{IP}\left(X;\mu_{1}\right)$ is plotted for $\mu_1=-10$ together with its Airy-kernel approximation in Fig.~\ref{fig:airy_kernel}. Good agreement is indeed observed at $X \! \simeq \! X_{e}=2^{1/4}\left|\mu_{1}\right|^{1/2} \! =3.760603093\dots$.

The analogous limiting behavior of the IP kernel \eqref{eq:K_IP_def} in the limit $-\mu_{1} \gg 1$ and $X\simeq X_e$ and $Y \simeq X_e$ is
\bea
&& \lim_{\mu_{1}\to-\infty}\!\frac{\mathcal{K}^{IP}\left(X_{e}-\frac{z}{2^{3/4}\left|\mu_{1}\right|^{1/6}},X_{e}-\frac{w}{2^{3/4}\left|\mu_{1}\right|^{1/6}};\mu_{1}\right)}{2^{3/4}\left|\mu_{1}\right|^{1/6}}  \nn\\ 
&&\qquad\qquad\qquad\qquad\qquad\qquad\qquad =K_{{\rm Ai}}\left(z,w\right).
\eea

\subsection{Asymptotics of $F_1^{IP}(0;\mu_1)$ as a function of $\mu_1$}\label{appendix:asymptotics_at_origin}

By putting $X=0$ in Eq. (\ref{eq:F_IP_bulk_approx2}), we obtain immediately the $\mu_1\gg1$ asymptotic of {$F_1^{IP}(0;\mu_1)$} given in Eq. (\ref{eq:F_IP_at_origin_bulk}).

To obtain the other limit $-\mu_1 \gg 1$, we can proceed in two alternative ways. The first corresponds to analysing the exact 
integral representation in Eq. (\ref{eq:F_IP_at_origin}) by choosing a vertical contour $\Gamma$ crossing the real axis at $\tau = \pi -\epsilon$, with $\epsilon \ll 1$. By substituting $\tau = (\pi - \epsilon) + i s$ with $s$ real, one can show that, to leading order, $F_1^{IP}(0;\mu_1) \sim e^{\pi \mu_1}$ for $\mu_1 \to -\infty$.  

There is an alternative, more physical, derivation of this asymptotic behavior when $\mu_1 \to - \infty$ which uses the WKB approximation. Indeed, plugging $x=x'=0$ into the eigenfunction representation of the kernel~(\ref{Kmu}), we obtain
\beq
\label{eq:F_at_origin}
N \rho_N\left(0\right)=K_{\mu}\left(0,0\right)=\sum_{k}\Theta\left(\mu-\epsilon_{k}\right)\left|\psi_{k}\left(0\right)\right|^{2}.
\eeq
At $-\mu_1\gg1$, we obtain the leading order result by keeping only the highest energy level $\epsilon_{\max}\simeq\mu$ in Eq.~(\ref{eq:F_at_origin}), and evaluating its corresponding eigenfunction using the WKB approximation.
In the parabolic approximation~(\ref{eq:quadratic_potential}) the classical turning points are given by $x_{\text{cl}}=\pm\sqrt{\left(V_{0}-\epsilon_{\max}\right)/V_{2}}$. Then the WKB approximation yields $\psi\left(0\right)\sim e^{-S}$ where
\bea
\label{eq:WKB_S}
S&=&\int_{-x_{\text{cl}}}^{0}\sqrt{\frac{2m}{\hbar^{2}}\left[V\left(x\right)-\epsilon_{\max}\right]}\,dx \nn \\
&=&\frac{\pi m^{1/2}\left(V_{0}-\epsilon_{\max}\right)}{2\sqrt{2}\:\hbar V_{2}^{1/2}}\simeq-\frac{\pi\mu_{1}}{2}.
\eea
Using Eq.~(\ref{eq:WKB_S}) in~(\ref{eq:F_at_origin}) and recalling Eq.~(\ref{eq:rho_scaling}), we arrive at the $-\mu_1\gg1$ asymptotic~(\ref{eq:F_IP_at_origin_WKB}) of $F_{1}^{IP}\left(0;\mu_{1}\right)$.

\section{Obtaining the {IP} kernel using eigenfunctions}
\label{appendix:eigenfunctions}

In this appendix, we obtain the kernel (and through it, the density) directly from its definition~(\ref{Kmu}), for a double-well potential $V(x)$ given in~(\ref{eq:potentialScaling}).
We begin by writing the time-independent Schr\"{o}dinger equation for the eigenfunctions $\psi_{k}\left(x\right)$ with the Hamiltonian~(\ref{ham}):
\beq
\label{eq:schrodinger_dim}
-\frac{\hbar^{2}\psi_{k}''\left(x\right)}{2m}+ V(x) \psi_{k}\left(x\right)=\epsilon_{k}\psi_{k}\left(x\right).
\eeq
If we now replace the exact double well potential $V(x)$, by the inverted parabola 
$V_0-V_2 x^2$, the spectrum becomes continuous. We can then
index the eigenfunctions by a continuous index $a$
defined as 
\beq
a=-\frac{m^{1/2}\left(\epsilon_k-V_{0}\right)}{2^{1/2}\hbar V_{2}^{1/2}} \, .\label{defa}
\eeq
It is convenient to look for solutions of \eqref{eq:schrodinger_dim} of the form
\beq
\label{eq:schrodinger_rescaling}
%\tilde{x}=\frac{2^{3/4}x}{w_{N}},\quad
\psi_{k}\left(x\right)=\frac{2^{3/8}}{w_{N}^{1/2}} \tilde \phi_{a}\left(\frac{2^{3/4}x}{w_{N}}\right).
\eeq
where $\tilde \phi_a(z)$ satisfies 
\beq
\label{eq:rescaled_schrodinger}
\tilde{\phi}_{a}''\left(z\right)+\frac{1}{4} z^{2}\tilde{\phi}_{a}\left(z\right)=a\tilde{\phi}_{a}\left(z\right),
\eeq
For each value of $a$ there are two linearly independent real solutions, 
$\tilde{\phi}_{a,+}\left(x\right)$ and $\tilde{\phi}_{a,-}\left(x\right)$,
given in Eqs.~(\ref{eq:phiplus}) and~(\ref{eq:phiminus}), which are respectively 
of even and odd parity. {They are}  orthonormalized in the following sense  (see Eq.~(6.14) in Ref.~\citep{Barton1986})
\bea
\label{eq:weird_delta_normalization}
&& \antiquad \int_{-\infty}^{\infty}\tilde{\phi}_{a,+}\left(z\right)\tilde{\phi}_{a',-}\left(z\right)dz=0 \\
&& \antiquad \int_{-\infty}^{\infty}\tilde{\phi}_{a,\pm}\left(z\right)\tilde{\phi}_{a',\pm}\left(z\right)dz=\delta\left(a-a'\right) {\pm}\frac{ \delta_{a,a'} }{4\cosh\left(\pi a\right)} \nn
\eea
where $\delta_{a,a'}$ is just the Kronecker delta function. 

It is natural to replace, in the large $N$ limit, the discrete sum in the exact formula for the kernel (\ref{Kmu}) 
by an integral over the spectral index $a$ of the inverted parabola. The integral is cutoff at the Fermi level
$\epsilon_k < \mu$ which corresponds to $a > - \mu_1$, where $\mu_1$ is given in \eqref{eq:mu1def}.
This yields to the following conjecture for the large $N$ limit of the kernel
$K_\mu\left(x,y\right)$ with the scaling form in the critical region given by Eq. \eqref{eq:K_scaling}
with the rescaled kernel given by Eq.~\eqref{eq:K_IP_eigenfunctions}
%Naftali: I commented out this equation because it appears in the main text already. In any case, if it is inserted, its label must be changed.
%\bea
%\label{eq:K_IP_eigenfunctions}
%&&\antiquad \mathcal{K}^{IP}\left(X,Y;\mu_{1}\right)=2\sqrt{\pi}\sum_{\sigma\in\left\{ +,-\right\} }\nn\\
%&&\qquad\qquad \int_{-\mu_{1}}^{\infty}\tilde{\phi}_{a,\sigma}\left(2^{3/4}X\right)\tilde{\phi}_{a,\sigma}\left(2^{3/4}Y\right)da,
%\eea
where we recall that $X=x/w_N$ and $Y=y/w_N$ are the rescaled coordinates in the critical
region. One can check, using Eqs. \eqref{eq:K_scaling}, \eqref{eq:K_IP_eigenfunctions} 
and \eqref{eq:weird_delta_normalization}, that this kernel is reproducible as defined
in Eq. (\ref{reprod}). 
Although the normalization in \eqref{eq:weird_delta_normalization} is somewhat unusual, 
the Kronecker delta part does not contribute when performing the continuous integral
in proving the reproducibility property (it contributes to a set of measure zero).

\section{Obtaining the {IP} kernel via the Green's function}
\label{Green's}

Here we show how the IP kernel can be derived using a Green's function method, which provides an alternative derivation  
of the expression for the kernel given in Eq. (\ref{kipd}). This derivation gives some complementary insight into the problem, in particular the emergence of the integral over a continuum of states in the final formula (\ref{kipd}).

We begin by taking the derivative of Eq.~\eqref{Kmu} with respect to $\mu$. This gives
%The zero temperature kernel is given by
%\begin{equation}
%K(x,y,\mu) = \sum_k^{\epsilon_k=\mu} \psi_k^*(x)\psi_k(y)= \sum_k\theta(\mu - \epsilon_k) \psi_k^*(x)\psi_k(y).
%\end{equation}
%and therefore
\begin{equation}
\partial_{\mu}K_{\mu}\left(x,y\right)=\sum_{k}\delta\left(\mu-\epsilon_{k}\right)\psi_{k}^{*}\left(x\right)\psi_{k}\left(y\right).
\end{equation}
We now recall 
\begin{equation}
\frac{1}{z-i 0^+}  = \pi i \delta(z) + P \frac{1}{z},
\end{equation}
where $P$ indicates that one should use the Cauchy principle part in any integrals,
and so
\begin{equation}
\delta(\mu-\epsilon) =\frac{1}{\pi}\rm {Im} \frac{1}{\mu -i 0^+-\epsilon} . 
\end{equation}
This leads to the key result
\begin{equation}
\label{eq:dK_dmu_and_Green}
\partial_{\mu}K_{\mu}\left(x,y\right)=\frac{1}{\pi}\text{Im}\,G_{\mu}\left(x,y\right),
\end{equation}
where $G_\mu$ is the Green's function obeying
\begin{equation}
\frac{\hbar^2}{2m}\frac{\partial^2}{\partial x^2} G_\mu\left(x,y\right) +[\mu - V(x)]G_\mu\left(x,y\right)=
\delta\left(x-y\right).
\end{equation}
Near $x=y=0$ we can use the Taylor expansion \eqref{eq:quadratic_potential}
 of the potential to write
\begin{equation}
\frac{\hbar^{2}}{2m}\frac{\partial^{2}}{\partial x^{2}}G_{\mu}\left(x,y\right)+\left(\mu-V_{0}+V_{2}x^{2}\right)G_{\mu}\left(x,y\right)=\delta\left(x-y\right).
\end{equation}
Now we make the change of variables $x=\alpha\zeta$ and $y=\alpha \zeta'$ (where $\alpha$ will be defined shortly), giving
\begin{eqnarray}
&&\frac{\partial^{2}}{\partial\zeta^{2}}G_{\mu}\left(\zeta,\zeta'\right)+\frac{2m\alpha^{2}}{\hbar^{2}}\left(\mu-V_{0}+V_{2}\alpha^{2}\zeta^{2}\right)G_{\mu}\left(\zeta,\zeta'\right) \nonumber\\
&& \qquad \qquad \qquad \qquad = \frac{2m \alpha}{\hbar^2}\delta(\zeta-\zeta')\,.
\end{eqnarray}
%We now choose $\alpha$ such that
%\begin{equation}
%\frac{2m \alpha^4}{\hbar^2}V_2 = \frac{1}{4},
%\end{equation}
%which gives 
%\begin{equation}
%\alpha = \left(\frac{\hbar^2}{8m V_2}\right)^{\frac{1}{4}} = \frac{w_N}{2^{\frac{3}{4}}},
%\end{equation}
%where $w_N$ is the length scale defined in Eq.~\eqref{eq:wNdef}.
%This means that we can write
%\begin{equation}
%G_\mu\left(\zeta,\zeta'\right) = \frac{2m \alpha}{\hbar^2}g_\mu\left(\zeta,\zeta'\right),
%\end{equation}
%with
Choosing
\begin{equation}
\alpha = \left(\frac{\hbar^2}{8m V_2}\right)^{1/4} = \frac{w_N}{2^{3/4}}\,,
\end{equation}
where $w_N$ is the length scale defined in Eq.~\eqref{eq:wNdef}, and defining
\begin{equation}
\label{eq:gmu_def}
g_{\mu}\left(\zeta,\zeta'\right)=\frac{\hbar^{2}}{2m\alpha}G_{\mu}\left(\zeta,\zeta'\right),
\end{equation}
we obtain the dimensionless equation
\begin{equation}
\frac{\partial^2}{\partial \zeta^2} g_\mu\left(\zeta,\zeta'\right) +\left(\frac{1}{4}\zeta^2 -a\right)g_\mu\left(\zeta,\zeta'\right)= \delta(\zeta-\zeta'),
\end{equation}
where
\begin{equation}
a= \frac{2m \alpha^2}{\hbar^2}(V_0-\mu) = -\mu_1
\end{equation}
is analogous to that introduced in Eq.~(\ref{defa}).

We now consider the homogeneous equation
\begin{equation}
\label{eq:Schrodinger_for_y_of_zeta}
\frac{\partial^2}{\partial \zeta^2} y(\zeta,a) +\left(\frac{1}{4}\zeta^2 -a\right)y(\zeta,a)=0 \;.
\end{equation}
The solutions of Eq.~\eqref{eq:Schrodinger_for_y_of_zeta} which match with the bulk or WKB solutions (which tend to zero for large argument due to the presence of  the term $-i 0^+$) are 
\begin{equation}
y(\zeta,a) = E^*\left(a,\zeta\right), 
\end{equation}
where $E\left(a,\zeta\right)$ is defined in the Handbook of Mathematical Functions 
in the chapter on parabolic cylinder functions \cite{abrom} (in the following, we use the notation of Ref.~\cite{abrom}). 
The $\zeta \to +\infty$ asymptotic of this function
 is
\begin{eqnarray}
&&\antiquad \antiquad E\left(a,\zeta\to +\infty \right)\simeq\sqrt{\frac{2}{\zeta}} \nn\\
&&\antiquad \times\exp\left\{ i\left[\frac{\zeta^{2}}{4}-a\ln(\zeta)+\frac{\text{arg}\left(\frac{1}{2}+ia\right)}{2}+\frac{\pi}{4}\right]\right\} ,
\end{eqnarray}
and thus clearly $y(\zeta,a)\to 0$ as $\zeta\to +\infty$. 
%For $\zeta\to-\infty$ we find
%\begin{eqnarray}
%&&E\left(a,\zeta\to-\infty\right)\simeq i\sqrt{\frac{2\left[1+\exp\left(2\pi a\right)\right]}{\left|\zeta\right|}} \times\nonumber \\
%&&\times\left(\exp\left\{ -i\left[\frac{1}{4}\zeta^{2}-a\ln\left(\left|\zeta\right|\right)+\frac{{\rm arg}\Gamma\left(\frac{1}{2}+ia\right)+\pi}{2}\right]\right\} \right. \nonumber \\
%&&-\left.\exp\left\{ \pi a+i\left[\frac{\zeta^{2}}{4}-a\ln\left(\left|\zeta\right|\right)+\frac{{\rm arg}\Gamma\left(\frac{1}{2}+ia\right)+\pi}{2}\right]\right\} \right). \nonumber\\
%\end{eqnarray}
%We thus see that this solution does not decay to zero as $\zeta\to-\infty$ (due to the second term). 
%However,
Moreover,  $y_-(\zeta,a)=E^*(a,-\zeta)$ is also a solution which has the correct behaviour as $\zeta\to -\infty$. 

It should also be noted that the complex solution can be written in terms of real functions as
\begin{equation}
\label{eq:E_and_W}
E\left(a,\zeta\right) = k^{-1/2} W\left(a,\zeta\right) + i k^{1/2} W(a,-\zeta),
\end{equation}
where $k=\sqrt{1+\exp(2\pi a)} -\exp(\pi a)$ and $k^{-1}=\sqrt{1+\exp(2\pi a)} +\exp(\pi a)$.  Here, $W\left(a,\zeta\right)$ and $W\left(a,-\zeta\right)$ are linearly independent solutions of Eq.~\eqref{eq:Schrodinger_for_y_of_zeta}
 with \cite{abrom}
\begin{equation}
\hat{W}\left[W\left(a,\zeta\right),W\left(a,-\zeta\right)\right]=1,
\end{equation}
where the Wronskian $\hat W[f,g]$ of two functions $f$ and $g$ is defined by $\hat W[f,g] = f'g-g'f$. 
From this we find that
\begin{equation}
\hat{W}\left[E^* \! \left(a,\zeta\right),E^* \! \left(a,-\zeta\right)\right]=k^{-1}+k.
\end{equation}
It is then straightforward to show that $g\left(\zeta,\zeta'\right)$ is given by
\begin{equation}
g\left(\zeta,\zeta'\right)=-\frac{E^* \! \left(a,\zeta\right)E^* \! \left(a,-\zeta'\right)}{k^{-1}+k},
\end{equation}
for $\zeta > \zeta'$, and the regime $\zeta < \zeta'$ is obtained through the symmetry $g\left(\zeta,\zeta'\right)=g\left(\zeta',\zeta\right)$.
Using this together with Eqs.~\eqref{eq:dK_dmu_and_Green}, \eqref{eq:gmu_def} and \eqref{eq:E_and_W} we find
\beq
\label{eq:Kprime_with_Ws}
\partial_{\mu}K_{\mu}\!\left(x,y\right)\!=\!\frac{2m\alpha}{\hbar^{2}}\frac{W\!\left(a,\zeta\right)\! W\!\left(a,\zeta'\right)\!+\!W\!\left(a,-\zeta\right)\! W\!\left(a,-\zeta'\right)}{\pi(k^{-1}+k)} .
\eeq
%\begin{eqnarray}
%K'(x,y,\mu)&=&\frac{2m \alpha}{\hbar^2}\frac{1}{\pi(k^{-1}+ k)}\nonumber\\
%& \times& [W\left(a,\zeta\right)W(a,\zeta') + W(a,-\zeta)W(a,-\zeta')]\nonumber\\
%\end{eqnarray}
We now integrate Eq.~\eqref{eq:Kprime_with_Ws} with respect to $\mu$ to obtain
\begin{eqnarray}
&& K_{\mu}\left(x,y\right) = \frac{2 m \alpha}{\pi \hbar^2 (k^{-1}+ k)} \int_{-\infty}^{\mu}\!d\mu' \nonumber \\ 
&& \left\{ W\left[\frac{2m\alpha^{2}\left(V_{0}-\mu'\right)}{\hbar^{2}},\zeta\right]W\left[\frac{2m\alpha^{2}\left(V_{0}-\mu'\right)}{\hbar^{2}},\zeta'\right]\right. \nonumber \\
&& + \left.W\left[\frac{2m\alpha^{2}\left(V_{0}-\mu'\right)}{\hbar^{2}},-\zeta\right]W\left[\frac{2m\alpha^{2}\left(V_{0}-\mu'\right)}{\hbar^{2}},-\zeta'\right]\right\} .\nn\\
\end{eqnarray}
Making the change of variables 
\begin{equation}
u'=\frac{2\left(V_{0}-\mu'\right)m\alpha^{2}}{\hbar^{2}}=\frac{\left(V_{0}-\mu'\right)\sqrt{m}}{\hbar\sqrt{2V_{2}}},\end{equation}
we obtain
\begin{eqnarray}
&&\antiquad\antiquad K_{\mu}\left(x,y\right) =\frac{1}{\alpha\pi (k^{-1}+ k)} \int_{-\mu_1}^\infty du' \nonumber \\
&&\antiquad \left[W\left(u',\zeta\right)W\left(u',\zeta'\right)+W\left(u',-\zeta\right)W\left(u',-\zeta'\right)\right].
\end{eqnarray}
%Making the change of variables 
%\begin{equation}
%u'=\frac{2\left(\mu'-V_{0}\right)m\alpha^{2}}{\hbar^{2}}=\frac{\left(\mu'-V_{0}\right)\sqrt{m}}{\hbar\sqrt{2V_{2}}},\end{equation}
%we obtain
%\begin{eqnarray}
%&&K_{\mu}\left(x,y\right) =\frac{1}{\alpha\pi (k^{-1}+ k)} \int_{-\infty}^u du' \nonumber \\
%&&   \left[W\left(-u',\zeta\right)W\left(-u',\zeta'\right)+W\left(-u',-\zeta\right)W\left(-u',-\zeta'\right)\right] \nonumber \\
%&& =\frac{1}{\alpha\pi (k^{-1}+ k)} \int_{-u}^\infty du' \nonumber \\ 
%&& \left[W\left(u',\zeta\right)W\left(u',\zeta'\right)+W\left(u',-\zeta\right)W\left(u',-\zeta'\right)\right].
%\end{eqnarray}
Introducing the sharp parity states defined in \cite{Barton1986} 
\begin{equation}
\psi_{\pm}\left(a,\zeta\right)=\frac{W\left(a,\zeta\right)\pm W\left(a,-\zeta\right)}{\sqrt{2}},
\end{equation}
we find
\begin{eqnarray}
&&K_{\mu}\left(x,y\right) =\frac{1}{2\pi \alpha \sqrt{1+\exp(2\pi a)}}\nonumber \\
&& \int_{-\mu_1}^{\infty}du'\left[\psi_{+}\left(u',\zeta\right)\psi_{+}\left(u',\zeta'\right)+\psi_{-}\left(u',\zeta\right)\psi_{-}\left(u',\zeta'\right)\right]. \nn\\
\label{gkf}
\end{eqnarray}
In \cite{Barton1986} the following inner product relations are demonstrated
\bea
&& \antiquad \int_{-\infty}^{\infty}{\psi}_{+}\left(a,z\right){\psi}_-\left(a',z\right)dz=0 \\
&& \antiquad \int_{-\infty}^{\infty}\psi_{\pm}\left(a,z\right)\psi_{\pm}\left(a',z\right)dz=2\pi \sqrt{1+\exp(2\pi a)} \nonumber \\ && 
\qquad \qquad \qquad \times \left[\delta\left(a-a'\right) {\pm}\frac{ \delta_{a,a'} }{4\cosh\left(\pi a\right)}\right] \;.
\eea
From these relations one can verify that the kernel \eqref{gkf} is reproducible. We note that the only difference
between $\psi_{\pm}(a,z)$ and $\tilde{\phi}_{a,\pm}(z)$ as defined in Eqs. (\ref{eq:phiplus}) and (\ref{eq:phiminus})  is their normalisation (see Eq. (\ref{eq:weird_delta_normalization})). Taking into account this and  the numerical difference between the length scales $\alpha$ and $w_N$ we see that the results in Eqs.~(\ref{kipd}) and (\ref{gkf}) indeed coincide.

{}


\begin{thebibliography}{}
\bibitem{footnote:mirror_symmetry}
Due to the mirror symmetry $x\leftrightarrow-x$ of the density, only the regime $x \ge 0$ is plotted in Figs.~\ref{fig:density} and \ref{fig:airy_kernel}.
\end{thebibliography}
\end{document}